\documentclass[11pt, a4paper,  notitlepage]{article}
\usepackage{amssymb}
\usepackage{graphicx}
\usepackage{amsmath}
\usepackage{amsfonts} 
\usepackage{pstricks} 
\usepackage{jheppub} 
\begin{document}
\title{
Exact oscillations and chaos on a non-Abelian coil 
}
\author[a,b]{Fabrizio Canfora}
\author[c,d]{Nicolas Grandi}
\author[e]{Marcelo Oyarzo} 
\author[e]{Julio Oliva}
\affiliation[a]{Facultad de Ingenier\'{\i}a, Arquitectura y Dise\~{n}o,  Universidad San Sebasti\'{a}n, sede Valdivia, \\ General Lagos 1163, Valdivia 5110693, Chile}
\affiliation[b]{\textit{Centro de Estudios Cient\'{\i}ficos (CECS), \\ Avenida Arturo Prat 514, Valdivia, Chile}}
\affiliation[c]{\textit{Departamento de Física, Universidad Nacional de La Plata, \\ CC67, 1900 La Plata, Argentina}}
\affiliation[d]{\textit{Instituto de Física La Plata, Consejo Nacional de Investigaciones Científicas y Técnicas\\ Diagonal 113 entre 63 y 64,  1900 La Plata, Argentina}}
\affiliation[e]{\textit{Departamento de Física, Universidad de Concepción, Casilla 160-C, Concepción, Chile }}

\emailAdd{fabrizio.canfora@uss.cl}
\emailAdd{grandi@fisica.unlp.edu.ar}
\emailAdd{juoliva@udec.cl}
\emailAdd{moyarzo2016@udec.cl}

\abstract{ 
We construct new exact solutions of the Georgi-Glashow model 
in $3+1$ dimensions. These configurations are periodic in time but  lead to a stationary energy density 
and no energy flux. Nevertheless, they possess a characteristic frequency which manifests itself through non-trivial resonances  
on
test fields.
This allows us to interpret them as non-Abelian self sustained coils. We show that for
larger
energies a transition to chaotic behavior takes place, which we characterize by Poincaré sections, Fourier spectra and exponential growth of the geodesic deviation in an effective Jacobi metric, the latter triggered by parametric resonances.
}
\maketitle
\section{Introduction}

Time-periodic configurations arising in nonlinear hyperbolic problems are notoriously difficult to construct (see \cite%
{time-periodicYM1,time-periodicYM2,time-periodicYM3} and references therein) and, at the same time, extremely interesting physically (see e.g. \cite{time-periodicYM4, time-periodicYM5,time-periodicYM6}). In Euclidean spaces,
the relevance of topologically non-trivial configurations which are periodic in Euclidean time, representing instantons at finite temperature, is particularly relevant for the analysis of the phase diagram of gauge theories \cite{BalaBook, manton}. 
%Even more, recently the
The interest in these configurations arises, in part, 
from the difficulty to study time dependent configurations in lattice gauge theories \cite{lattice1, lattice2}.
It also 
%The interest in this type of configurations 
experienced a remarkable growth 
in the recent years,
due to the intensive research in out-of-equilibrium physics (see e.g. \cite{entanglement5, entanglement6, entanglement7, entanglement8, entanglement9, entanglement9a1, entanglement9a2, entanglement9b, entanglement9c, entanglement9d, entanglement10, entanglement11, entanglement12} and references therein).

In the present 
%work
paper we %will
construct new exact, time dependent solutions %of 
to the Yang-Mills-Higgs system in $3+1$ dimensions, with quite intriguing physical properties. These configurations are periodic in real time in such a way that the energy-density is stationary and
%, moreover, the
their
non-Abelian Poynting vector vanishes, so that there is no energy flux. In spite of 
this,
%these properties,
as we will show below,
they 
%these configurations 
possess a characteristic frequency which manifests itself through non-trivial resonances of test fields, charged under the non-Abelian gauge symmetry, which propagate in these 
backgrounds.
%background configurations. 
These new analytic solutions possess genuine non-Abelian features as 
they
%these 
can be interpreted as non-Abelian self-sustained coils. 

Besides the intrinsic interest to construct analytical time-dependent configurations, the technical tools allow to discuss very interesting open 
%question in
questions on
the chaotic behavior of Yang-Mills theory. The analysis of
chaos
%the chaotic behavior 
in non-Abelian gauge theories raised huge interest since the early years soon after the discovery of Yang-Mills theory (see \cite{chaosYM1, chaosYM2, chaosYM3, chaosYM4, chaosYM5} and references therein). In recent years, two references in particular \cite{chaosYM5.1, chaosYM5.2} triggered a burst of 
%interest 
activity
on this topic due to the discovery of novel relations with holography
and 
quantum chaos
%and so on
(see \cite{chaosYM6, chaosYM7, chaosYM8, chaosYM9} and references therein). The usual starting point of these analyses is a
homogeneous Ansatz for the Yang-Mills-Higgs fields with, very often, the Higgs field in the fundamental representation, which only depend on time, in such a manner that the corresponding field equations can be analyzed with the available tools of chaotic dynamics (see \cite{chaosYMreview}). On the other hand, this starting point prevents, in many situations, to include non-trivial topological fluxes, that either need some non-trivial
dependence on space-like coordinates, or the presence of the Higgs field in the adjoint representation, in order to get a gauge-invariant version of the magnetic flux. Therefore, if one is interested in the analysis of the interplay of topology and chaos, it is important to generalize a little bit the notion of homogeneous field and to construct an Ansatz 
in which
%in such a way that
the fields depend
non-trivially
%in a non-trivial way also
on the spatial coordinates, keeping alive the topological fluxes, but in such a way that the field equations reduce to a dynamical system. 

An important technical tool  
to succeed in the aforementioned
%underlying the present
construction 
%has been
turns out to be
the non-spherical hedgehog Ansatz developed for the Skyrme model, originally introduced in \cite{56a0}-\cite{crystal4}, that allowed to discover the first analytic and topologically non-trivial solutions in the Skyrme model which are periodic in time in such a way that the energy-momentum tensor is static \cite{56a, 56a1}. As explained below, in a certain sense the 
%present
results 
presented here
represent an extension of those in \cite{56a} and \cite{56a1} to the Yang-Mills-Higgs case, with the Higgs in the adjoint representation of the gauge group.

At a first glance, the analytic solutions representing non-Abelian self-sustained coils, to be described in the following sections, could suggest the appearance of some integrable sector of the theory. In fact, this is not the case: the chaotic behavior appears anyway. However, in the analysis of the chaotic 
regime,
%behavior,
the analytic solutions manifest themselves through ``integrability islands" in the corresponding Poincaré sections. One of the main tools that we will use in the analysis of chaotic dynamics 
was
%is the one
introduced in \cite{cerruti-pettini}
and
%which 
is based on the Jacobi metric \cite{arnold}. Our analysis shows that such a tool, which to the best of our knowledge has not been employed so far in the analysis of chaos in Yang-Mills theory, is actually
very effective when compared with different techniques.

The paper is organized as follows: in Section \ref{sec:model} the conventions and the Georgi-Glashow model are presented. In Section \ref{sec:ansatz} we 
%present 
introduce
the time dependent Ansatz for the Yang-Mills and Higgs fields in flat spacetime. Later, 
the new exact solutions of the system are derived, as well as some of their perturbations. The cases with and without vacuum expectation value are studied separately
in sections \ref{sec:solutions.vev} and \ref{sec:solutions.novev} respectively.
In Section 5 we study the resonance frequencies of the configurations with a quantum scalar field probe in the fundamental of $SU(2)$. Some remarks and conclusions are given in the last section.

\section{Basic setup}
\label{sec:setup}
In this section, the model and the time-dependent Ansatz are introduced, together with the corresponding equations of motion and the resulting energy momentum tensor and non-Abelian Poynting vector.
\subsection{The model}
\label{sec:model}
Our starting point is the Georgi-Glashow model for $SU\left(2\right)$, with field content given by a Lie algebra valued $1$-form gauge potential $A$ and a Higgs field $\Phi$  which transforms in the adjoint representation. They are algebra valued objects
\begin{equation}
A=A_{\ \mu }^{a}t_{a}dx^{\mu }\ ,\qquad \qquad \Phi =\Phi ^{a}t_{a}\ ,
\label{eq:model.fields}
\end{equation}
where we consider anti-Hermitian matrices $t_{a}\equiv i\sigma _{a}$, where $\left\{ \sigma _{a}\ ,\ a=1,2,3\right\} $ are the Pauli matrices. These generators fulfill $t_{a}t_{b}=-\delta _{ab}-\varepsilon _{abc}t_c$.

The action 
for the model 
reads
\begin{equation}
I\left[ A,\Phi \right] =
\int d^{4}x\sqrt{-g}
\left( 
	-\frac{1}{4e^2}F^{a \mu\nu }F_{a \mu \nu }
	-\frac{1}{2e^2}D_{\mu }\Phi^{a} D^{\mu }\Phi_{a}
-\frac{\lambda}{4}  \left(\Phi^a \Phi_a -\nu^2 \right) ^{2} \right)
\ ,
\label{eq:model.action}
\end{equation}
where $e$ is a positive gauge coupling constant, $\lambda$ is a positive scalar self coupling, and $\nu$ is the vacuum expectation value of the Higgs field. As usual, the field strength and the covariant derivative are defined by
\begin{eqnarray}
F_{\mu \nu } &=&\partial _{\mu }A_{\nu }-\partial _{\nu }A_{\mu }+\left[A_{\mu },A_{\nu }\right] \ ,  
\label{eq:model.gaugecurvature} \\
D_{\mu }\cdot &=&\nabla _{\mu }\cdot +\left[ A_{\mu },\cdot \right] \ .
\label{eq:model.covariantderivative}
\end{eqnarray}

The field equations are obtained by computing the stationary variation with respect to the fields $A_{\ \mu }^{a}$ and $\Phi ^{a}$ which respectively give the following expressions
\begin{eqnarray}
D_{\mu }F^{\mu \nu }-\left[ \Phi ,D^{\nu }\Phi \right] &=&0\ , 
\label{eq:model.equations.YM}
\\
D_{\mu }D^{\mu }\Phi -e^2 \lambda \left( \Phi^a\Phi_a -\nu ^{2}\right) \Phi
&=&0\ .
\label{eq:model.equations.Higgs}
\end{eqnarray}

The energy momentum tensor of this model is computed by varying the action with respect to the metric, resulting in 
\begin{equation}
T_{\mu \nu }=T_{\mu \nu }^{\sf Gauge}+T_{\mu \nu }^{\sf Higgs}\ ,
\label{eq:model.energymomentumtensor}
\end{equation}
with
\begin{align}
&T_{\mu \nu }^{\sf Gauge}=\frac{1}{e^{2}}\left( 
	F_{a \mu \lambda }F_{\ \nu }^{a\ \lambda}-\frac{1}{4}g_{\mu \nu}F^{a \rho \sigma }F_{a \rho \sigma }
\right)  \ ,
\label{eq:model.energymomentumtensor.YM}
\\
&T_{\mu \nu }^{\sf Higgs}=\frac{1}{e^{2}}
\left(
D_{\mu }\Phi^{a} D_{\nu }\Phi_{a} 
-\frac{1}{2}g_{\mu\nu}D_{\sigma
}\Phi^{a} D^{\sigma }\Phi_{a} 
-g_{\mu \nu }  \frac{\lambda e^2}{4}\left( \Phi^{a} \Phi_{a}-\nu^{2}\right) ^{2}  \right) \ .  
\label{eq:model.energymomentumtensor.Higgs}
\end{align}
From now on we set $e=1$ without loss of generality, since the only relevant combination is $\lambda e^2$.
\subsection{The time dependent Ansatz}
\label{sec:ansatz}
In the present section we define an appropriate Ansatz which allows us to solve the field equations analytically with a time dependent profile. 

Let us first fix the geometry  considering flat spacetime in cylindric coordinates
\begin{equation}
ds^{2}=-dt^{2}+dz^{2}+d\rho ^{2}+\rho ^{2}d\varphi ^{2}\ .  \label{eq:ansatz.metric}
\end{equation}
The range of the coordinates are the usual,
%as usual
$\varphi \in \left[ 0,2\pi \right] $ with $\varphi \sim \varphi +2\pi \ ,\ \rho \in \lbrack 0,+\infty \lbrack $ and $t,z\in \mathbb{R\ }$. 
In this background we define our Ansatz for the gauge field and Higgs fields, as
\begin{eqnarray}
A &=&-\frac{W( t) }{\sqrt{2}}\left(t_{1}\, \rho d\varphi-t_{2}\,d\rho \right) -\frac{1}{2}t_{3}\,d\varphi \ ,  \label{eq:ansatz.YM} \\
\Phi  &=&G( t)\, t_{3}\ .  \label{eq:ansatz.Higgs}
\end{eqnarray}
Both the gauge $W(t)$ and Higgs $G(t)$ profiles depend explicitly on time.
%{This setup will allow us to describe a self-sustained non-Abelian coil, where the functions $W(t)$ and $G(t) $ are periodic functions fixed by the field equations (as it will be discussed in a moment).}
%
The non-Abelian field strength defined in (\ref{eq:model.gaugecurvature}) for the Ansatz (\ref{eq:ansatz.YM}) reads 
\begin{equation}
F=\frac{\dot{W}}{\sqrt{2}}\left(\,dt\wedge d\rho \,t_{2}-\rho \,dt\wedge d\varphi \,t_{1}\right)-W^{2}\rho \,d\rho \wedge d\varphi \,t_{3}\ .
\label{eq:ansatz.F}
\end{equation}
It has two electric components, one of them is along the second generator of the gauge group while pointing in the radial spatial direction, while the other is aligned with the 
first
%second
generator and it points around the cylinder. The magnetic field is aligned with the third generator and it
goes
%is pointing 
along the axis of the cylinder. 

\bigskip

With the above Ansatz, the energy momentum tensor has a natural cylindrical symmetry, and it can be  written as
\begin{equation}
T_{\mu \nu}\,dx^{\mu }\otimes dx^{\nu }=\frac{1}{e^2}\left({\cal E}\,dt^2-p_\perp (d\rho^2+\rho^2d\theta^2)-p_z\,dz^2\right)\,,
\label{eq:ansatz.energymometumtensor}
\end{equation}
with
\begin{align}
&\mathcal{E}= \frac{1}{2}\left( \dot{G}^{2}+%
\dot{W}^{2}\right)+\frac12W^2\left(4G^{2}+W^{2}\right)+\frac{%
\lambda }{4}\left( G^{2}-\nu ^{2}\right) ^{2}  \ ,  \label{eq:ansatz.energy}
\\
&p_\perp=-\frac12\left(\dot{G}^{2}+  W^4\right)+\frac{\lambda }{4}\left( G^{2}-\nu ^{2}\right)
^{2}\,,
\label{eq:ansatz.pperp}
\\
&p_z=p_\perp+W^{2}\left(2G^2+W^2\right)-\frac12 W'^2\,.
\label{eq:ansatz.pz}
\end{align}
It is worth to emphasize that, in spite of considering a time dependent configuration, there are no energy fluxes.
%, which
This feature
can be interpreted as an 
interplay
%equilibrium
between the non-Abelian character of the solution and the time dependence of the gauge fields, in such a way that the trace in the definition of the energy momentum tensor cancels out the radiation of the gauge field. We will discuss this feature in more detail in section \ref{sec:resonance}. 

\newpage 

To give some physical content to the above construction, let us first recall one of the most
%One of the
useful features of the Georgi-Glashow model:
%is that
the presence of a scalar field in the adjoint representation allows
us
to construct a gauge invariant quantity representing the effective Abelian gauge field of the theory
\begin{equation}
{F}_{\sf eff}={\rm tr}\left( \Phi F\right) \ .
\label{eq:ansatz.Feef}
\end{equation}
For the configuration (\ref{eq:ansatz.YM}) and (\ref{eq:ansatz.Higgs}), the above projection gives 
\begin{equation}
{F}_{\sf eff}=\rho\, G\,W^{2}d\rho \wedge d\varphi
\ .  \label{eq:ansatz.Feff.components}
\end{equation}
In the present case, the $2$-form (\ref{eq:ansatz.Feff.components}) corresponds to an effective uniform Abelian magnetic flux along the $z$-axis. The 
exact
configurations that will be discussed in the following are periodic in time, hence the effective Abelian magnetic field will be periodic as well. 

Now
let us consider a cylinder of radius $R_{0}$ inside of which the fields are %non-vanishing and 
given by the Ansatz (\ref{eq:ansatz.YM})-(\ref{eq:ansatz.Higgs}), while 
they vanish 
outside.
%the cylinder the fields are zero.
In order to match the fields in the interior of the cylinder with 
those
%the fields 
outside
it,
we require the usual 
Maxwell
junction conditions for the
corresponding 
Abelian part 
(\ref{eq:ansatz.Feff.components}).
%of the field defined in Eq. (\ref{efective F}) 
These conditions tell us that the normal component to the interface of the effective Abelian magnetic field must be continuous,
which is satisfied by $B_{\sf{eff}}=G\left( t\right) W\left( t\right) ^{2}\partial _{z}$. Also, %as we show below 
since
the Poynting vector is zero everywhere,
%hence
there is no energy flux outside the cylinder. Consequently, if we are able to construct explicitly exact solutions for the gauge and Higgs profiles which are periodic in time, then such configurations can be interpreted as coils with a self-generated $AC$ current.

\bigskip

\begin{figure}
\begin{center}
\includegraphics[width=0.48\textwidth]{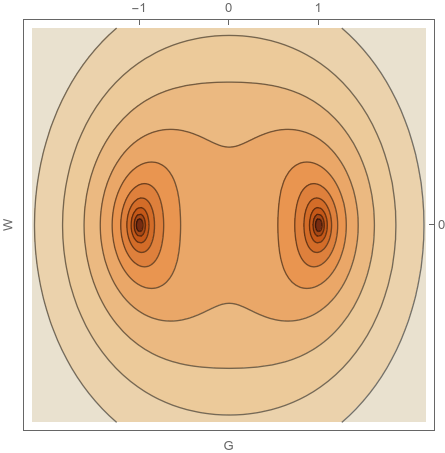}
\ 
\includegraphics[width=0.48\textwidth]{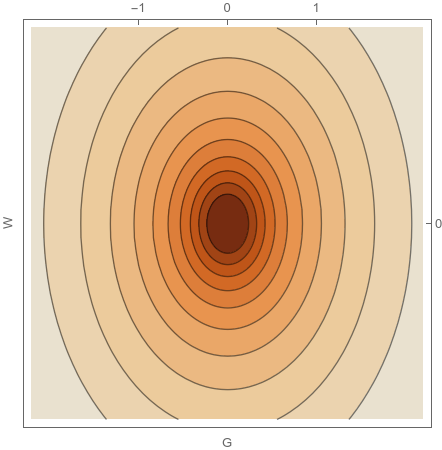}
\end{center} 
\vspace{-0.5cm}
\caption{
\label{fig:levelcurves} 
Level curves of the effective potential $V$ for the cases $\nu\neq0$ (left) and $\nu=0$ (right).}
\end{figure}

A very important property of the ansatz for the gauge and Higgs fields 
given in Eqs. (\ref{eq:ansatz.YM}) and (\ref{eq:ansatz.Higgs}) 
is that it reduces the full coupled system of non-linear partial differential equations %corresponding to the Georgi-Glashow model 
to the following two coupled ordinary differential equations
    \begin{align}
    &\frac{d^{2}G}{dt^{2}}+4W^{2}G +\lambda G\left( G^{2}-\nu ^{2}\right)=0\ ,
    \label{eq:ansatz.eom.Higgs} \\
    &\frac{d^{2}W}{dt^{2}}+4G^{2}W +2W^{3}=0\ .  \label{eq:ansatz.eom.YM}
    \end{align}
It is conceptually useful to rewrite this system of second order differential equations as a Newtonian system for the time-dependent variables $\left( G  ,W  \right)$ 
in the form
%as
%
\begin{equation}
\frac{d^{2}G}{dt^{2}} =-\frac{\partial }{\partial G}V\left(
G,W\right) \ ,\qquad \qquad\frac{d^{2}W}{dt^{2}}=-\frac{\partial }{\partial W}V\left( G,W\right) \ ,  
\label{eq:ansatz.newtoninan}
\end{equation}
in terms of the effective potential
\begin{equation}
V( G,W)  =
2W^{2}G^{2}+\frac{1}{2}W^{4}+\frac{\lambda }{4}\left(G^{2}-\nu ^{2}\right)^2\ .  \label{eq:ansatz.potential}
\end{equation}
The configuration with non-trivial expectation value has zero vacuum energy thanks to the additive constant $\lambda \nu^4/4$. This potential is bounded from below and has two global minimum at $G=\pm \nu \ ,\ W=0$ and a saddle point at $G=0,W=0$. A plot of the level curves of this potential is shown in Fig. \ref{fig:levelcurves}.

As a first interesting result notice that, integrating
%Integrating 
the system (\ref{eq:ansatz.newtoninan})-(\ref{eq:ansatz.potential}) once, we 
%get
recover
the conservation of the energy density ${\cal E}$, in spite of the fact that the field configuration is time dependent. This is consistent with the absence of energy fluxes in our configuration. 

Naively, one could conclude that this configuration is static and hence there is no a characteristic frequency of the system. Nevertheless, this is not the case as we will show  in section \ref{sec:resonance} by computing the time-dependent transition amplitude of a scalar probe field in the adjoint representation propagating in the exact solutions of the above form. Such transition amplitude discloses a clear resonance effect when the frequency of the test field matches the characteristic frequency of the background solutions. The present situation is reminiscent of the spin-from-isospin effect for Skyrmions and non-Abelian monopoles \cite{spinfromisospin1, spinfromisospin2, spinfromisospin3} in which case the energy-momentum tensor is spherically symmetric and yet these configurations are not spherically symmetric in the obvious sense as the angular momentum operator is naturally supplemented by an extra term arising from the internal symmetry group. This fact is behind the commonly used statement ``gauge field are invariant up to an internal transformation''. 

Notice that the equations \eqref{eq:ansatz.eom.Higgs} and \eqref{eq:ansatz.eom.YM} have the shift symmetry
\begin{equation}
t\to t-t_0\ ,
\label{eq:ansatz.shift}
\end{equation}
which implies that one of the integration constants of the system sets the zero of the time variable. Moreover, they have the scaling invariance
\begin{align}
&(t, W,G,\nu,\lambda) \rightarrow \left(\frac tT ,TW, TG, T\nu,\lambda \right)\ ,
\label{eq:ansatz.scaling}
\end{align}
where $T$ is an arbitrary constant. For vanishing $\nu$ this implies that a second integration constant sets the time scale and the overall scale of the fields. For $\nu$ finite, these can be fixed by the value of $\nu$.

\bigskip
 
It is worth emphasizing that in a vast majority of papers in the available literature on chaos in Yang-Mills theory, the non-Abelian gauge potential and the Higgs field are chosen to be homogeneous in space, so that they only show a non-trivial dependence on the time variable $A_{\mu }=A_{\mu }(t)$, $\Phi =\Phi (t)$. We will refer to this type of configurations as ``trivially homogeneous'', since for them both the field equations as well as the physical observables (such as the energy-density and the effective magnetic flux) only depend on time. Even if this dependence could appear to be too restrictive, it is actually justified in situations where the time gradients are much larger than the space ones ({\em cf.} \cite{chaosYM1}-\cite{chaosYMreview}). 
This allows to use the well known techniques and ideas from the theory of chaos in dynamical systems, like for example those presented on \cite{cerruti-pettini-cohen} and \cite{cerruti-pettini}.
For this reason, since the early days of chaos in Yang-Mills theory ({\em cf.} \cite{ChaosNew1}) until the more recent references on this topics ({\em cf.} \cite{ChaosNew2} and references therein), trivially homogeneous configurations are considered. %as they give rise to effective Hamiltonians %(which very often look like interacting quartic oscillators) 
%whose chaotic features can be analyzed using sound and well-understood tools. 

One of the contributions of the present work is to extend the notion of homogeneous fields in order to apply chaos theory to more general Yang-Mills-Higgs configurations. In particular, % our definition of ``homogeneous configurations'' is weaker, since
we define a configuration as ``homogeneous'' if and only if the Ansatz describing it reduces the complete set of Yang-Mills-Higgs field equations to a consistent dynamical system of second order autonomous ordinary differential equations, for purely time dependent unknown functions. Obviously, a ``trivially homogeneous'' configuration is also homogeneous in our sense, but the converse is not true. Indeed, the explicit example above is not homogeneous in space, as both the gauge potential and the energy-momentum tensor depend on space-like coordinates. However, such dependence has been chosen in such a way that the field equations realize a dynamical system of two second order autonomous ODE's in two unknown time dependent functions.

This technical result allows to extend considerably the range of applicability of chaos tools to gauge theory, keeping alive both the genuine non-Abelian character of the configurations
as well as the non-Abelian magnetic flux.

\newpage 

\section{Exact solutions}
\label{sec:solutions}
In this section we present our exact solutions and analyze their properties, studing separately the cases with a without vacuum expectation value. In each case, we explore the vacuum and perturbative solutions, the pure Yang-Mills and pure Higgs cases, and the solutions with both fields turned on.

\subsection{Configurations with non-vanishing vacuum expectation value}
\label{sec:solutions.vev}
In this section we will consider configurations with non-vanishing vacuum expectation value $\nu\neq 0$. 

\paragraph{Perturbative solution:}
The first trivial observation in 
%In
this case
is that 
there is a vacuum static solution in which $W(t)=0$ and $G(t)=\pm \nu$. Such solution can be perturbed as
\begin{align}
&W(t)=\epsilon w(t)\ ,\\
&G(t)=\pm \nu+\epsilon g(t)\ ,
\end{align}
where $\epsilon$ is a small parameter and $w(t)$ and $g(t)$ are new unknown functions. Plugging this back into the equations of motion and expanding to first order in $\epsilon$, we get a perturbative solution
\begin{align}
&W(t)=\epsilon \sin(2\nu(t-t_0)+\delta)\ ,\\
&G(t)=\pm\nu+\epsilon a \cos(\sqrt{2\lambda}\nu(t-t_0))\ ,
\end{align}
where $\epsilon$ now becomes a small integration constant, and $a,t_0$ and $\delta$ are integration constants of order one. 
Notice that these solutions are periodic only when $\sqrt{\lambda/2}=p/q$ with   $p,q\in\mathbb{N}$. The period then reads
\begin{equation}
    t\sim t+\frac\pi\nu\sqrt{\frac2\lambda}\,p=t+\frac{\pi}{\nu}q
\end{equation}

\paragraph{Pure Yang-Mills solution:}
There is a pure Yang-Mills sector of the theory, which is obtained setting $G(t)=0$. In this case, the field equations (\ref{eq:ansatz.eom.Higgs})-(\ref{eq:ansatz.eom.YM}) reduce to the equations of a quartic oscillator, namely
\begin{equation}
\frac{d^{2}W}{dt^{2}}+2W^{3}=0\ ,
\end{equation}
that can be solved in the form
\begin{equation}
W( t) =\pm a\ {\rm sn}\!\left( a\left( t-t_{0}\right) ,-1\right) \ ,
\label{eq:solution.pureYM.vev}
\end{equation}
where ${\rm sn}(x,m)$ is the Jacobi elliptic sine function, and $a$ is a constant of integration. Notice that the same constant sets both the time scale and the amplitude of the oscillation. This can be traced back to the scaling symmetry \eqref{eq:ansatz.scaling}, taking into account that the value of $\nu$ does not enter into the present pure Yang-Mills solution.

We can calculate the energy density of the configuration according to the expression \eqref{eq:ansatz.energy}, obtaining
\begin{equation}
{\cal E}=\frac{1}{4}\left(2a^4+\lambda \nu^4\right)
 \ ,
\label{eq:energy.pureYM.vev}
\end{equation}
where we see that the energy density is conserved. 

Solutions \eqref{eq:solution.pureYM.vev} are periodic, their period can be obtained from the periodicity properties of the Jacobi elliptic sine, resulting in the expression
\begin{equation}
t\sim t+\frac{2}aK_{20}(-1) \ ,
\label{eq:period.pureYM.vev}
\end{equation}
where the function $K_{pq}(m)$ has been defined according to
\begin{eqnarray}
    K_{pq}(m)=p\,K(m)+i\,q\,K(1-m)\ .
    \label{eq:Kpq}
\end{eqnarray}
In this expression, $K$ is the complete elliptic integral of the first kind, and $p,q\in\mathbb{N}$. Here and in what follows, we are choosing the values of $p$ and $q$ as the smallest integers that make the resulting period real.
\bigskip

\paragraph{Pure Higgs solution:}
There is also a pure Higgs configuration, which is obtained by setting $W(t)=0$ and solving the remaining equation for $G(t)$, resulting in
\begin{equation}
G(t)=\pm\nu \sqrt{\frac{2q}{1+q}}\,{\rm sn}\!\left(\sqrt{\frac{-\lambda }{1+q}}\,\nu\,(t-t_0),q\right) \ ,
\label{eq:solution.pureHiggs.vev}
\end{equation}
where  $q$ is a constant of integration.

As for the pure Yang-Mills case, this is a periodic solution whose period is given by that of the Jacobi sine, in the form 
\begin{equation}
    t\sim t+\frac{2}{\nu}\sqrt{\frac{1+q}{-\lambda}} K_{20}(q)\ ,
    \label{eq:period.pureHiggs.vev}
\end{equation}
where $K_{pq}$ defined as in equation \eqref{eq:Kpq}.

Solution \eqref{eq:solution.pureHiggs.vev} is explicitly real for $q<-1$. However, using the definition of and properties of the Jacobi elliptic functions, it can be analytically continued to $q\in(-1,0]$ in the form
\begin{equation}
G(t)=\pm\nu \sqrt{\frac{-2q}{1+q}}\,{\rm sc\!}\left(-\sqrt{\frac{\lambda }{1+q}}\,\nu\,(t-t_0),1-q\right) 
\ ,
\label{eq:solution.pureHiggs.vev.2}
\end{equation}
where $\mbox{sc}(x,m)=i\,\mbox{sn}(-ix,1-m)$ is another Jacobi function. 

Expression \eqref{eq:solution.pureHiggs.vev.2} is again periodic, but in this case the period is written in the form
\begin{equation}
    t\sim +\frac{2}{\nu}\sqrt{\frac{1+q}{\lambda}} K_{22}(1-q)\ ,
    \label{eq:period.pureHiggs.vev.2}
\end{equation}
which connects smoothly to \eqref{eq:period.pureHiggs.vev} as $q\to-1$.

This configuration has an energy density given by
\begin{equation}
    {\cal E}=\frac\lambda 4\left(\frac{1-q}{1+q}\right)^2\nu^4\ ,
\end{equation}
which is again conserved.

\bigskip

\paragraph{Solution with both fields:}
For the generic case with non-vanishing Higgs, the solution reads
\begin{eqnarray}
G( t) &=&\pm_1\sqrt 2\,\nu\,{\rm dn}\!\left( \sqrt{8-\lambda}\,\nu \left( t-t_{0}\right), \frac{\lambda}{8-\lambda}
\right) \ , \label{eq:solution.bothG.vev}\\
W( t) &=&
\pm_2
\sqrt{\frac{\lambda(\lambda-4)}{8-\lambda}}\,\nu\,{\rm sn}\!\left( \sqrt{8-\lambda}\,\nu \left( t-t_{0}\right), \frac{\lambda}{8-\lambda}
\right)  \ ,  \label{eq:solution.bothW.vev}
\end{eqnarray}
where ${\rm dn}^2(x,m)=1-m\,{\rm sn}^2(x,m)$ is another Jacobi elliptic function.  This solution is explicitly real for $\lambda \in [4,8)$. There is no integration constant controlling the frequency of the oscillation, nor its amplitude. However, the vacuum expectation value parameter $\nu$ changes the amplitude and the frequency of the configuration in the same amount, due to the scaling symmetry discussed in the previous section  \eqref{eq:ansatz.scaling}. The period is given by

\begin{equation}
    t\sim t+\frac{2}{\nu\sqrt{8-\lambda}} K_{22}\!\left( \frac{\lambda }{8-\lambda }\right)
    \,.
    \label{period0}
\end{equation}
 
Using the identities and the relations between the Jacobi elliptic functions one can write \eqref{eq:solution.bothG.vev}-\eqref{eq:solution.bothW.vev} 
in an alternative form which is 
manifestly real for $\lambda>8$. In such case we have
\begin{eqnarray}
G\left( t\right)  &=&\pm_1\sqrt{2}\,\nu\, \text{dc}\!\left(  \sqrt{\lambda -8}\,\nu\,\left(
t-t_{0}\right) ,1-\frac{\lambda }{8-\lambda }\right) \ , \\
W\left( t\right)  &=&\pm_2  \sqrt{\frac{\lambda \left( \lambda -4\right) }{\lambda -8}}\,\nu\,\text{sc}\!\left(\sqrt{\lambda -8} \,\nu \left( t-t_{0}\right) ,1-%
\frac{\lambda }{8-\lambda }\right) \ .
\end{eqnarray}
Where $\mbox{dc}(x,m)=\mbox{dn}(-i x,1-m)$ is a further elliptic function. The period now reads

\begin{equation}
    t\sim t+\frac{2}{\nu\sqrt{\lambda-8}}K_{22}\!\left(1-\frac{\lambda }{8-\lambda}\right)\ .
    \label{period1}
\end{equation}

The $\lambda=8$ can be integrated from the equations (\ref{eq:ansatz.eom.Higgs})-(\ref{eq:ansatz.eom.YM}) and it reads
\begin{eqnarray}
G\left( t\right) &=&\sqrt 2\,\nu\, \sin \! \left( 2\sqrt{2}\nu \left( t-t_{0}\right)
\right) \ , \\
W\left( t\right) &=&2\,\nu\, \cos \!\left( 2\sqrt{2}\nu \left(
t-t_{0}\right) \right) \ .  \label{W(t) vev}
\end{eqnarray}
Here, $t_{0}$ is the only integration constant of the solution.
The period of this solution can be written as
\begin{equation}
t\sim t+\frac{\pi}{\sqrt2\nu}\ .
\end{equation}

For any value of the coupling $\lambda$, the energy density of this exact configuration is given by the  expression
\begin{equation}
{\cal E}=\frac{1}{4}(2\lambda-7) \lambda\nu^4\  , \\
\end{equation}
which behaves smoothly in the $\lambda\to8$ limit.

\begin{figure}
\begin{center}
\includegraphics[width=0.325\textwidth]{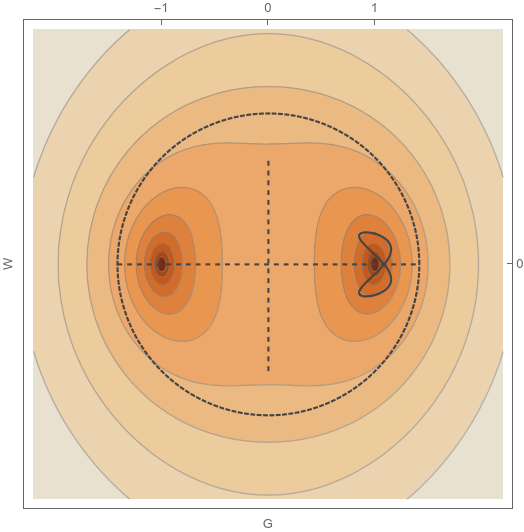}
%\hspace{.001\textwidth} 
\includegraphics[width=0.325\textwidth]{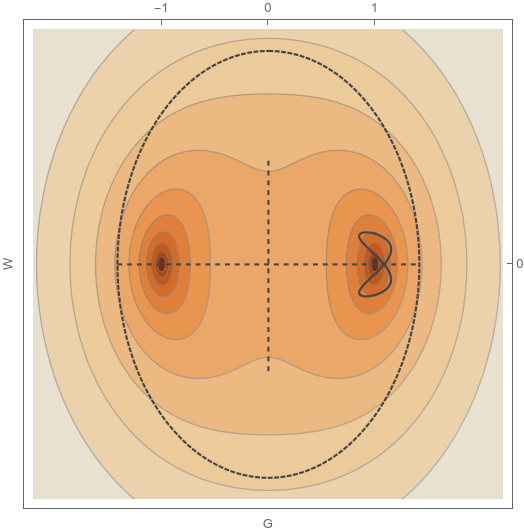}
%\hspace{.001\textwidth} 
\includegraphics[width=0.325\textwidth]{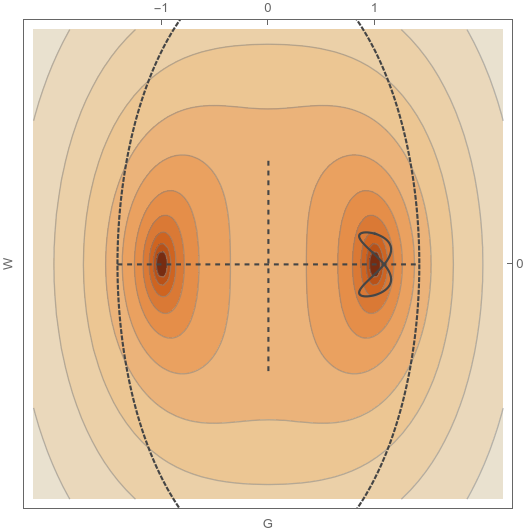}
\end{center}
\vspace{-.5cm}
\caption{\label{potential2}Solutions with non-vanishing vacuum expectation value $\nu\neq0$, for the particular cases $\lambda=6,8,12$ from left to right. The continuous gray line is the perturbative solution, while the dotted line is the exact one. The vertical and horizontal dashed lines represent the pure Yang-Mills and  pure Higgs solutions respectively, the last corresponding to the smallest amplitude $q\to-\infty$.}
\end{figure}

In Fig.\ref{potential2} we overlap the solutions we found in this section with the level curves of the effective potential (\ref{eq:ansatz.potential}) for $\nu =1, \lambda=6,8,12$.

\newpage
\subsection{Configurations with vanishing vacuum expectation value}
\label{sec:solutions.novev}

\paragraph{Perturbative solution:}
With $\nu =0$ we still have a static solution, now at $W(t)=G(t)=0$, that can perturbed to obtain
\begin{align}
&W(t)=\epsilon (t-t_0)\ ,\\
&G(t)=\epsilon  a(t-t_0)\ .
\end{align}
Higher order perturbations result in further corrections to the overall coefficient of the linear term, up to order $\epsilon^3$ at which there is an additional correction which goes as $(t-t_0)^5$.

\paragraph{Pure Yang-Mills solution:}
The pure Yang-Mills configuration is the same as in the case with non-vanishing vacuum expectation value, which is to be expected since the Higgs field plays no role in it.

\paragraph{Pure Higgs solution:}
Regarding the pure Higgs configuration, it satisfies the equation of motion
\begin{equation}
\frac{d^{2}G}{dt^{2}}+ \lambda G^{3}=0\ .
\end{equation}
This is again a quartic oscillator, with solution
\begin{equation}
    G(t)=\pm \sqrt2a\,{\rm sn}\left(a\sqrt\lambda(t-t_0),-1\right)\ ,
    \label{eq:solution.pureHiggs.novev}
\end{equation}
where $a$ is an integration constant. 
The period takes the form
\begin{equation}
    t\sim t+\frac 2{a\sqrt\lambda}K_{20}(-1)\ .
    \label{eq:period.pureHiggs.novev}
\end{equation}
The energy density on the other hand, reads
\begin{eqnarray}
    {\cal E}=a^4\lambda \ .
    \label{eq:energy.pureHiggs.novev}
\end{eqnarray}
It is interesting to notice that formulas (\ref{eq:solution.pureHiggs.novev}) to (\ref{eq:energy.pureHiggs.novev}) can be obtained from the corresponding equations for the finite vacuum expectation value case, by taking the limit $\nu\to0$ and $q\to-1$ with the constraint $\nu/\sqrt{1+q}=i \,a$.

\begin{figure}[ht]
\begin{center}
\includegraphics[width=0.325\textwidth]{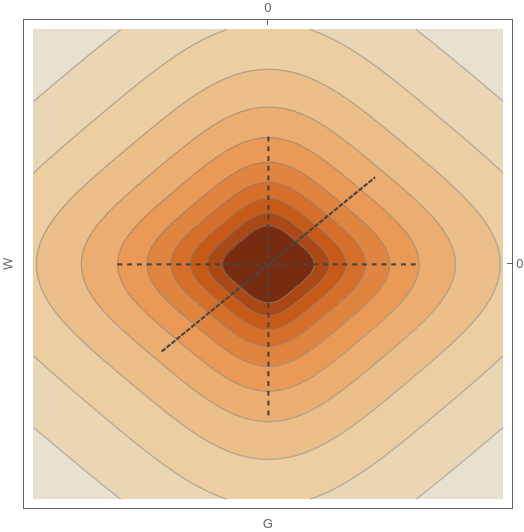}
%\hspace{.01\textwidth} 
\includegraphics[width=0.319\textwidth]{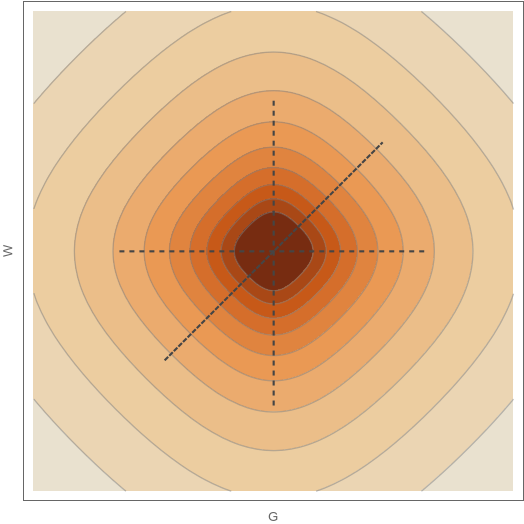}
%\hspace{.01\textwidth} 
\includegraphics[width=0.325\textwidth]{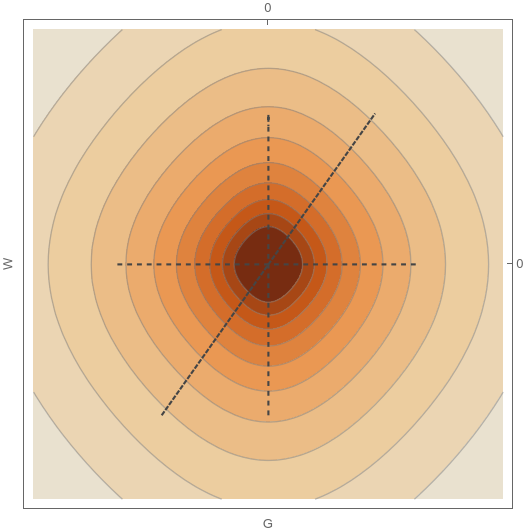}
\end{center}
\vspace{-.5cm}
\caption{\label{potential3}Solutions with non-vanishing vacuum expectation value $\nu=0$, for the particular cases $\lambda=1,2,3$ from left to right. In orange is the exact solution. The vertical and horizontal lines represent the pure Yang-Mills and  pure Higgs solutions respectively.}
\end{figure}
\paragraph{Solution with both fields:}
The fact that for the linearly perturbed solution we have a Higgs profile $G$ that is proportional to the Yang-Mills profile $W$, suggests that in the non-perturbative case we can try to reduce the equations (\ref{eq:ansatz.eom.Higgs})-(\ref{eq:ansatz.eom.YM}) into a unique equation, by considering the ansatz
\begin{equation}
G( t) =\pm \sqrt{\frac{2}{4-\lambda }}W( t) \ .
\label{G W}
\end{equation}
Here the proportionality factor has been chosen so that the resulting equations for $G(t)$ and $W(t)$ coincide. Notice that the shape of the potential provides that $\lambda $ must be positive and the equation (\ref{G W}) implies that $\lambda <4\ $. The resulting master equation is given by
\begin{equation}
\frac{d^{2}W}{dt^{2}}+2\left( \frac{8-\lambda}{4-\lambda }\right)  W^{3}=0\ ,
\end{equation}
which can be solved by
\begin{equation}
W( t) =\pm a\
 {\rm sn}\!\left( a\sqrt{\frac{8-\lambda }{4-\lambda}} \left(
t-t_{0}\right) ,-1\right) \ , \label{eq:solution.bothW.novev}
\end{equation}
%
%where ${\rm sn}$ is the Jacobi sine function.  
Notice that as before the amplitude $a$ is tied to the frequency due to the scaling symmetry, but now it is an integration constant. Consequently, the profile for the Higgs field reads
\begin{equation}
G( t) =\pm \sqrt {\frac{2}{4-\lambda }}\,a\,{\rm sn}\!\left( a\sqrt{\frac{8-\lambda}{4-\lambda}}\left( t-t_{0}\right) ,-1\right) \ . \label{eq:solution.bothG.novev}
\end{equation}
These solutions are explicitly real for $\lambda<4$ and cannot be extended to $\lambda>4$. 

The energy density of this configuration is
\begin{equation}
{\cal E}=\frac{a^{4}\left( \lambda -8\right) \left( \lambda -6\right) }{2\left( \lambda
-4\right) ^{2}}\  ,
\label{energy.vanishing.vev}
\end{equation}
while the period can be obtained in terms of the complete elliptic integral of the first kind % $K$ %(see the Appendix I) 
%as
%
\begin{equation}
t\sim t+\frac{2}{a}\sqrt{\frac{\lambda -4}{\lambda -8}}\,K_{20}( -1) \ .
\end{equation}
In Fig.\ref{potential3} the solutions are shown, together with  a level plot of the effective potential (\ref{eq:ansatz.potential}). % for $\nu =0, \lambda=6,8,12$.
\newpage

\section{Chaotic behaviour}
At a first glance, one could think that the appearance of the nice analytic solutions described in the previous sections may hint at the integrability of the Yang-Mills-Higgs sector described by the Ansatz in Eqs. (\ref{eq:ansatz.YM}) and (\ref{eq:ansatz.Higgs}). This possibility becomes quite clear taking into account that using the homogeneous Ansatz which is usually employed in the analysis of chaos in Yang-Mills theory (see \cite{chaosYM6, chaosYM7, chaosYM8, chaosYM9}\ and references therein) it has not been possible to find analytic solutions, to the best of our knowledge. In the following sections we will show that this is not the case: the chaotic behavior appears nevertheless, if one increases the energy of the system. 

To characterize the chaotic regime, we will use three different and somewhat complementary techniques:
\begin{enumerate}
    \item {\bf Poincaré sections:} The phase space of the system is $4$-dimensional and can be parameterized by the coordinates $(G,W)$ and the canonical momenta $(p_G=\dot G, p_W=\dot W)$. The conservation of the energy \eqref{eq:ansatz.energy} reduces in one the dimensionality of the space where the trajectories develop. Poincaré sections are then constructed by performing one further projection onto the plane $(W,p_W)$.

    Regular trajectories appear in the Poincaré section as sets of points that can be connected with smooth curves. Chaotic behavior on the other hand, corresponds to sparse sets that fill the section. 
    
    The
    presence of analytic solutions manifests itself through ``integrability islands''.

    \item {\bf Fourier analysis:}  Chaos can 
    often
    be confused with a quasiperiodic behavior, a combination of linear oscillators with non-commensurable frequencies. In order to exclude the latter possibility of our analysis, % below 
    we consider the discrete Fourier spectrum of one of the canonical variables. 

    A non-smooth Fourier spectrum is a clear signature of a chaotic regime. 

    \item {\bf Geodesic divergence:}
    In classical mechanics, the time evolution of a Newtonian system of the kind defined by (\ref{eq:ansatz.newtoninan})-(\ref{eq:ansatz.potential}) can be described as  a non-affine parametrization of the geodesic curves on a manifold endowed with the so-called Jacobi metric \cite{arnold, loris}, defined according to
\begin{align}
 ds^2=g_{ij} \,dq^idq^j  &=2({\cal E}-V)\left(dW^2+dG^2\right) \; ,
 \nonumber\\
 &=4({\cal E}-V)^2\,dt^2\ ,
\end{align}
where $i,j$ run on the independent generalized coordinates $q^i=(G,W)$ and ${\cal E}$ is the energy of the system \eqref{eq:ansatz.energy}.  

The relation between the curvature of the manifold and the stability of the geodesics is expressed %mathematically 
in terms of the Jacobi-Levi-Civita equation for the Jacobi field $\eta ^{i}$, measuring the deviation 
%of
between
two infinitesimally close geodesics
%associated to a geodesic with tangent vector $v^{j}$%
%
\begin{equation}
{\nabla_s^{2}\eta ^{i}}-\mathcal{R}^{i}{}
_{jkl}\frac{dq^j}{ds}\frac{dq^k}{ds}\eta^{l}=0\ . \label{JLC eq}
\end{equation}
where $\nabla_s$ is the covariant derivative.  
In a two dimensional manifold the Riemann tensor can be written in terms of the scalar curvature $\mathcal{R}$ in the form ${\cal R}^i{}_{jkl}={\cal R}(\delta^i{}_{k}g_{jl} -\delta^i{}_{l}g_{jk})/2$.
This implies that 
\begin{equation}
{\nabla^2_s\eta ^{i}}+\frac{\mathcal{R}}2\left(\eta^i-\frac{dq^i}{ds}\frac{dq^j}{ds}\eta_j
\right)=0\ . \label{eq}
\end{equation}
Where in the second term we used the fact that $s$ is an affine parameter and the tangent vector is normalized to one. Contracting with $dq_i/ds$ and $\varepsilon_{ij}\,dq^j/ds$ (with $\varepsilon_{ij}$ the Levi-Civita tensor) and taking into account the geodesic equation $\nabla_s(dq^i/ds)=0$, we can write
\begin{equation}
{\frac{d^2\eta_\perp}{ds^2}+\frac{\mathcal{R}}2\eta_\perp
=0\ ,\qquad\qquad\quad
\nabla_s^{2}\eta_\parallel} =0\ , 
\end{equation}
Here we have defined $\eta_\parallel=\eta^i(dq^i/ds)$ and $\eta_\perp=\epsilon_{ij}\eta^i(dq^j/ds)$.

It is clear that a negative scalar curvature $\mathcal{R}<0$ would lead to solutions with an exponential grow in time for $\eta_\perp$. % _{\mathbf{1}}$. 
However, for %a 
our
Newtonian system the Ricci scalar $\mathcal{R}$ %for a generic potential 
is given by
\begin{align}
\mathcal{R}&=
\frac{1}{2\left( {\cal E}-V\right)^3 }
\left[
4W^2\left(2G^2+W^2\right)^2
+
G^2\left(4W^2+\lambda\left(
G^2-\nu^2\right)
\right)^2
\right]+
\nonumber\\&~~~~
+
\frac{1}{2\left( {\cal E}-V\right)^2 }
\left[
(4+3\lambda)G^2+10W^2-\lambda\nu^2
\right]
\ ,
\end{align}
Recalling that ${\cal E}-V>0$, the curvature scalar has a chance to be negative only when $G$ 
%$q_{1}$ 
and $W$ 
%$q_{2}$ 
are small enough,
so that the last term in the second line superseeds the rest. 
In most %of the
configurations this is not the case, thus the instabilities we eventually find %below for the system (\ref{system eta1 eta2}), 
should come from parametric 
resonance, as the scalar ${\cal R}$ is time dependent (for a similar situation, see \cite{cerruti-pettini}).

The
equation for the nearby geodesic deviation
%reduces to 
can be rewritten as
\begin{equation}
\label{eq para Y}
\frac{d^{2}Y}{dt^{2}}+\Sigma\,  Y =0\ ,
\end{equation}
where $Y=\eta_\perp/\sqrt{{\cal E}-V}$ and $\Sigma$ is a function of time defined as
\begin{eqnarray}
\Sigma  &=&
2{\cal R}({\cal E}-V)^2
-\frac{1}{2({\cal E}-V)}\frac{d^2V}{dt^2}-\frac{3}{4({\cal E}-V)^2}\left(\frac{dV}{dt}\right)^2
\end{eqnarray}
The form of the solution $Y$ as a function of 
time
%the trajectory $\left( q_{1}\left( t\right) ,q_{2}\left( t\right) \right) $ in the phase space 
gives us insight about the behaviour of the perturbation of the configuration, varying the initial conditions. If $Y$ is constant, it gives a signal of stability, while if $Y$ grows exponentially in time, the system could develop a chaotic behavior in that region. 
\end{enumerate}

\newpage
\subsection{Chaos with vanishing vacuum expectation value} 
The solution in \eqref{eq:solution.bothW.novev}-\eqref{eq:solution.bothG.novev} provides a uni-parametric family of analytic solutions with parameter $a$, in terms of which the energy is fixed through \eqref{energy.vanishing.vev}. By evaluating the solution at the initial time $t_0$, we obtain a set of initial conditions with periodic evolution. 

To depart from the analytic solution, we write one of the canonical variables in terms of the energy, say $\dot G^2(t_0)=2({\cal E}-V)-\dot W(t_0)$, and then move the value of the energy ${\cal E}$ away from \eqref{energy.vanishing.vev}. An insightful way to parameterize the remaining freedom in in terms of the initial magnetic flux, which according to \eqref{eq:ansatz.Feff.components} is proportional to $G(0)W(0)^2$, and the initial deviation of the configuration from a  purely Abelian one, that in terms of \eqref{eq:ansatz.F} can be identified as proportional to $\dot W(0)$.

\begin{figure}[hb]
\vspace{.3cm}
\begin{center}
\includegraphics[width=1.0\textwidth]{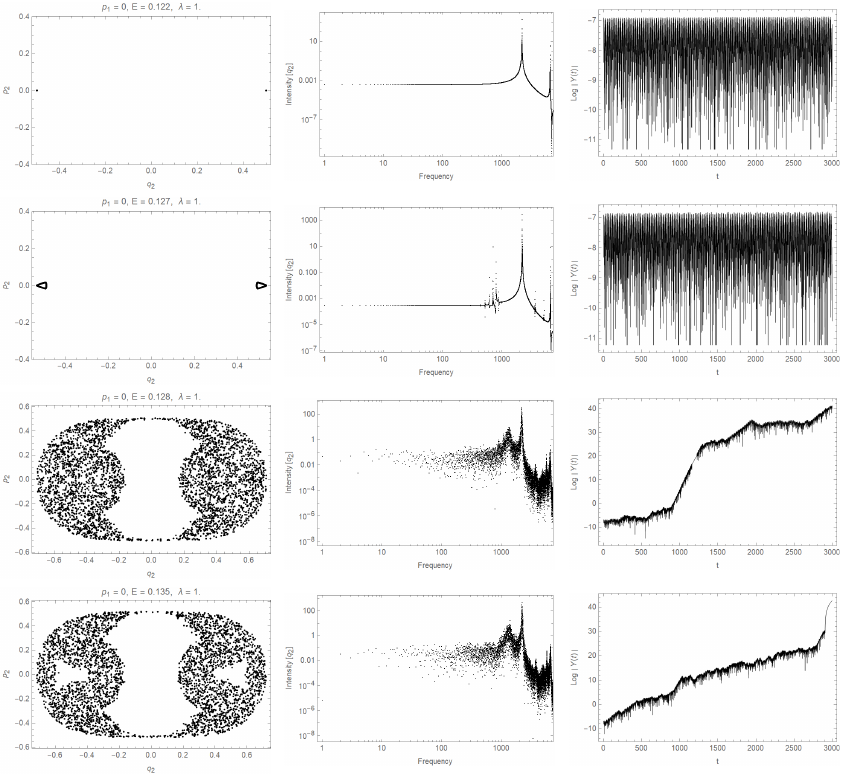}
\end{center}
\vspace{-.5cm}
\caption{Transition to chaos with vanishing vacuum expectation value. From left to right: Poincaré section, frequency spectrum and the logarithm of the geodesic deviation $\log|Y|$. We evolved up to $t_f=15000$ with $a=0.5$. Notice that between ${\cal E}=0.127$ and ${\cal E}=0.128$ there is a transition to chaos.}
\label{fig:plots chaos sin vev}
\end{figure}

\newpage

\begin{figure}
\begin{center}
\includegraphics[width=1.0\textwidth]{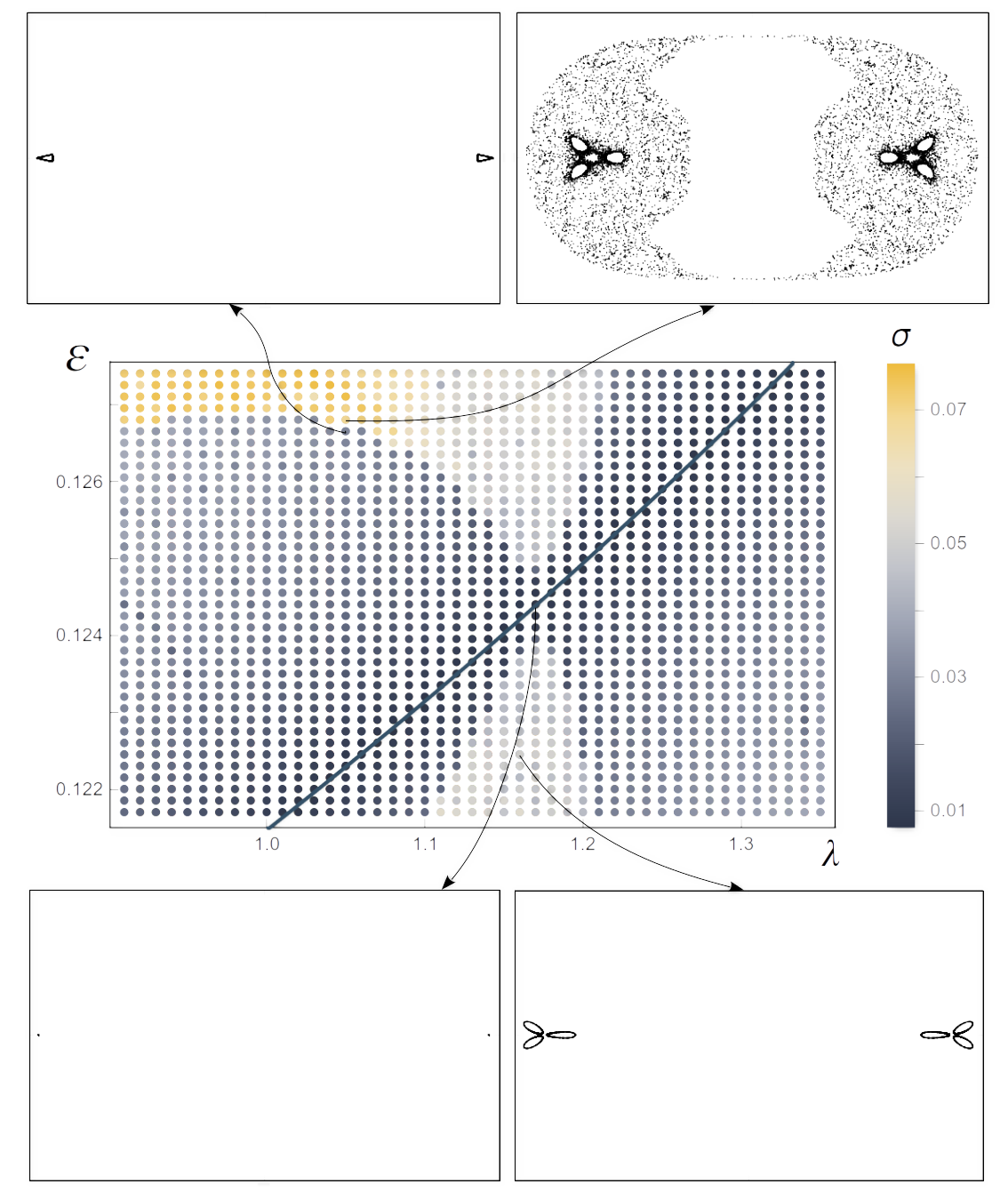}
\end{center}
\caption{Phase diagram in the energy ${\cal E}$ versus coupling $\lambda$ plane,   with vanishing vacuum expectation value. The initial magnetic flux vanishes $G(0)W(0)^2=0$ and the initial deviation from a purely Abelian configuration is $\dot W(0)=0.3818$. At each point, the chaotic nature of the solution is quantified by the mean quadratic dispersion of the points on the Poincaré section corresponding to the $\dot W$ versus $W$  plane. Some of such sections are depicted in the in-plots. The diagonal line that crosses the diagram correspond to the locus of the exact solution \eqref{eq:solution.bothW.novev}-\eqref{eq:solution.bothG.novev}.}
\label{fig:phase.diagram.novev}
\end{figure}

As we increase the energy with a fixed value of the coupling $\lambda$, we find chaotic behaviour above a critical value, which can be identified in the Poincaré sections, in the spectrum, and in the geodesic deviation, see the Fig. \ref{fig:plots chaos sin vev}. 

The critical energy for the transition to chaos is minimal for an intermediate value of the coupling $\lambda$, at which the chaotic behaviour also shows up at small energies, see Fig.\ref{fig:phase.diagram.novev}. This sets a narrow ``bridge'' between two ``islands'' of regular behavior, which is crossed by the locus of the exact solutions, as depicted by the blue regions of Fig.\ref{fig:phase.diagram.novev}. 

\begin{figure}
\begin{center}
\includegraphics[width=1.0\textwidth]{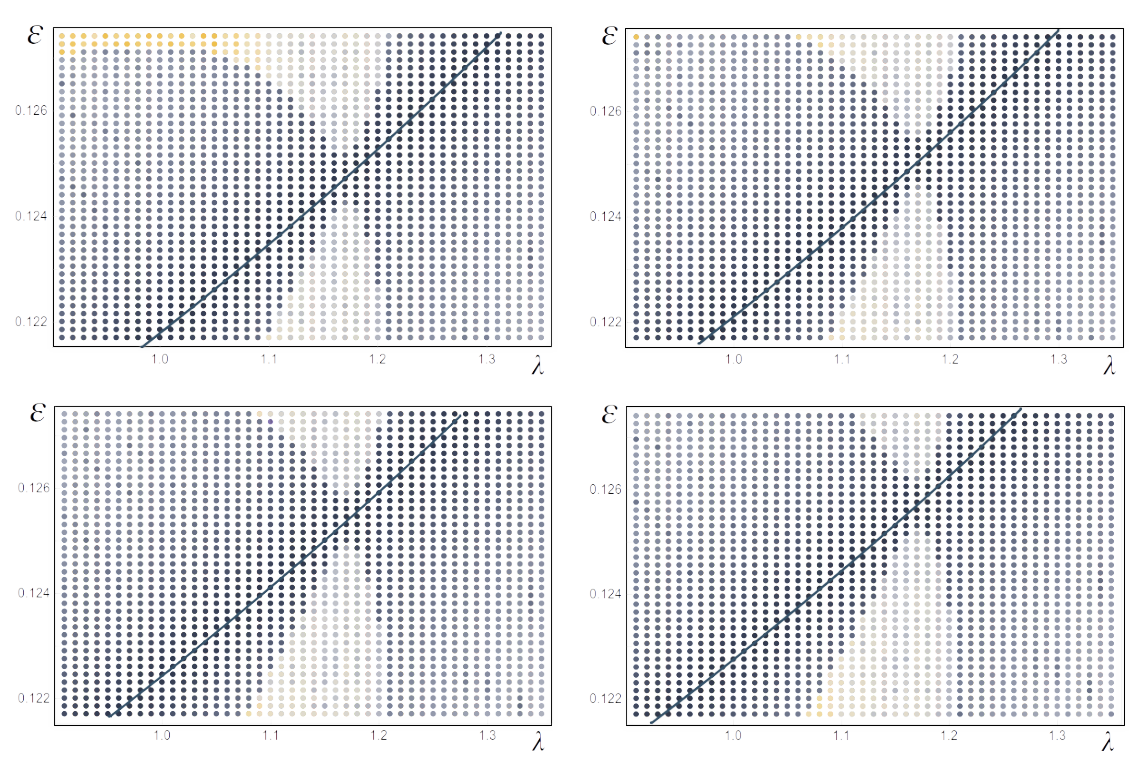}
\end{center}
\vspace{-.5cm}
\caption{Phase diagrams in the energy ${\cal E}$ versus coupling $\lambda$ plane, with vanishing vacuum expectation value, for different values of the initial deviation from a purely Abelian configuration $\dot W(0)$. The top row corresponds to $\dot W(0)=0.3823$ and $\dot W(0)=0.3828$ from left to right, while the bottom row has $\dot W(0)=0.3833$ and $\dot W(0)=0.3838$ respectively.  The initial magnetic flux vanishes $G(0) W(0)^2=0$. The loci of the exact solutions are depicted by the solid line.}
\label{fig:phase.diagram.NA.novev}
\end{figure}

\begin{figure}
\begin{center}
\includegraphics[width=1.0\textwidth]{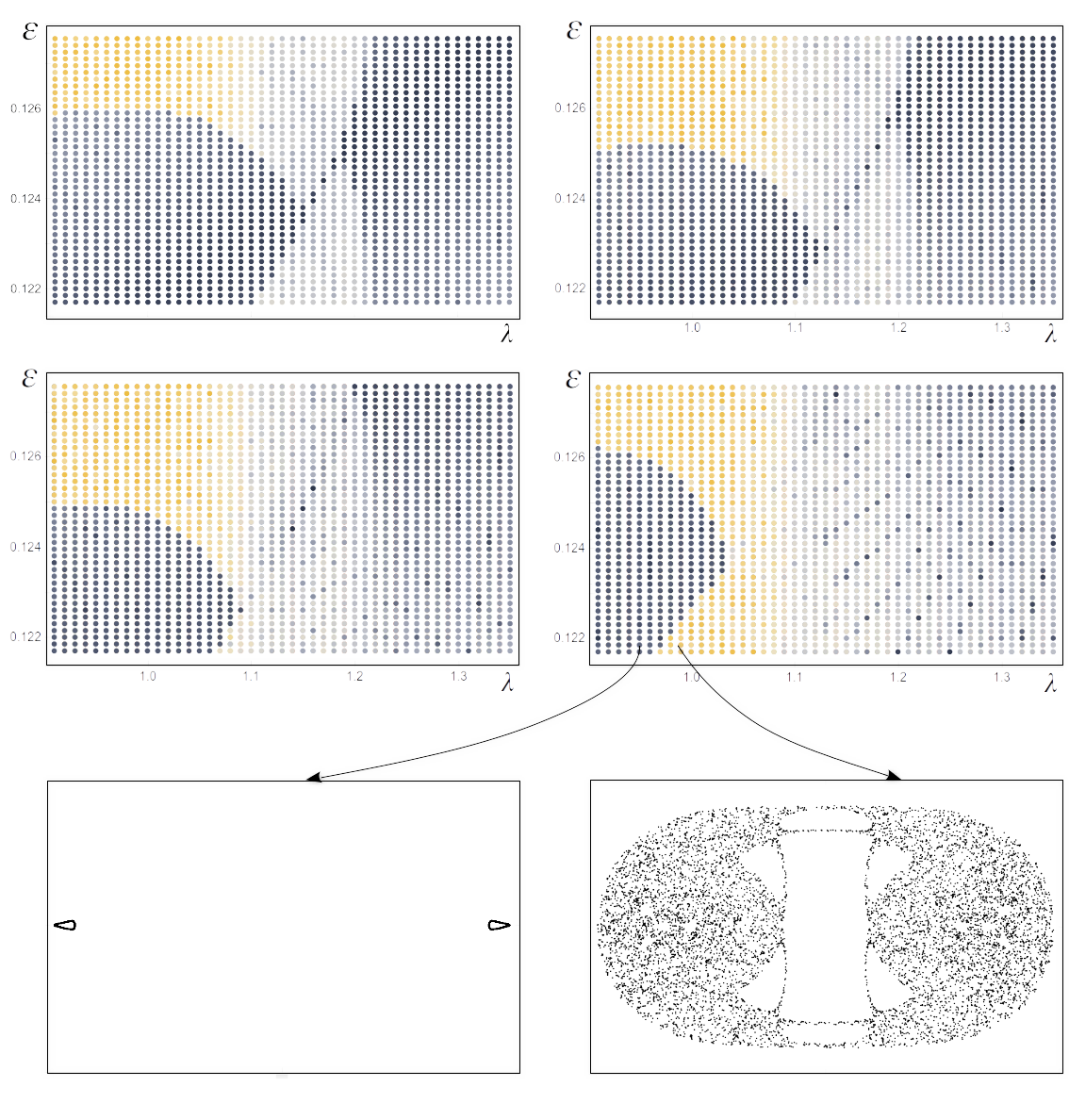}
\end{center}
\caption{Phase diagrams in the energy ${\cal E}$ versus coupling $\lambda$ plane, with  vanishing expectation value, for different values of the initial magnetic flux $G(0)W(0)^2$. The top row corresponds to $G(0)W(0)^2=(0.05)^3$ and $G(0)W(0)^2=1$ from left to right, while the middle row has $G(0)W(0)^2=(1.5)^3$ and $G(0)W(0)^2=2^3$ respectively.    The initial deviation from a purely Abelian configuration is $\dot W(0)=0.3818$. The last row shows two Poincaré sections in the $W$ versus $\dot W$ plane, corresponding to the non-chaotic and chaotic regimes of the last phase diagram. Even if there are no exact solutions for these values of the parameters, for small flux there is a ``regularity bridge'' joining the two regular regions of the phase diagram, which dissipates as the flux grows.}
\label{fig:phase.diagram.flux.novev}
\end{figure}

Finally, the regularity islands at large and small coupling get wider as the initial deviation from a purely Abelian configuration grows, see Fig.\ref{fig:phase.diagram.NA.novev}. On the other hand, as the initial flux gets larger, the exact solutions disappear, but the structure of two regularity islands joined by a bridge persists for a while before dissipating into chaos, see Fig. \ref{fig:phase.diagram.flux.novev}.

~

\newpage

\subsection{Chaos with non-vanishing vacuum expectation value}

The solution \eqref{eq:solution.bothG.vev}-\eqref{eq:solution.bothW.vev} is devoid of integration constants. We will proceed as before in order to explore the phase space, evaluating the initial conditions using the solution and deforming one of the fields away from the analytic regime by varying the energy ${\cal E}$, and charactherizing the remaining freedom in terms of the initial magnetic flux $G(0)W(0)^2$ and the initial deviation from a purely Abelian configuration $\dot W(0)$.
\begin{figure}[hb]
\vspace{.3cm}
\begin{center}
\includegraphics[width=1.0\textwidth]{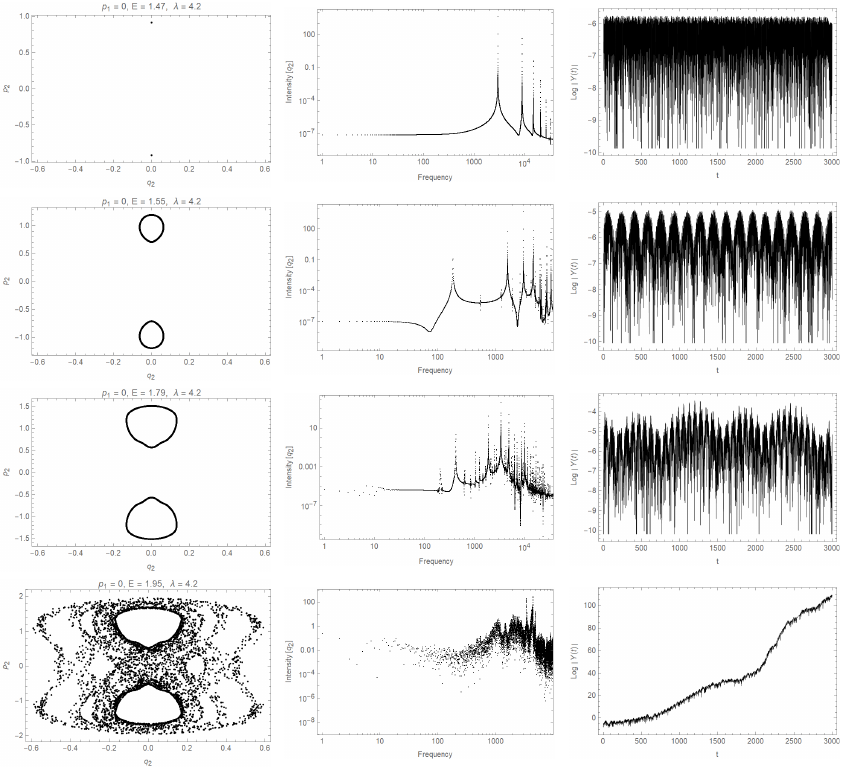}
\end{center}
\caption{Transition to chaos with non-vanishing vacuum expectation value. From left to right: Poincaré section, frequency spectrum and the logarithm of the geodesic deviation $\log|Y|$. We evolved up to $t_f=15000$ with $\lambda=4.2$. Notice that between ${\cal E}=1.195$ and ${\cal E}=0.179$ there is a transition to chaos.}
\label{fig:plots chaos con vev}
\end{figure}

\newpage

\begin{figure}
\begin{center}
\includegraphics[width=1.0\textwidth]{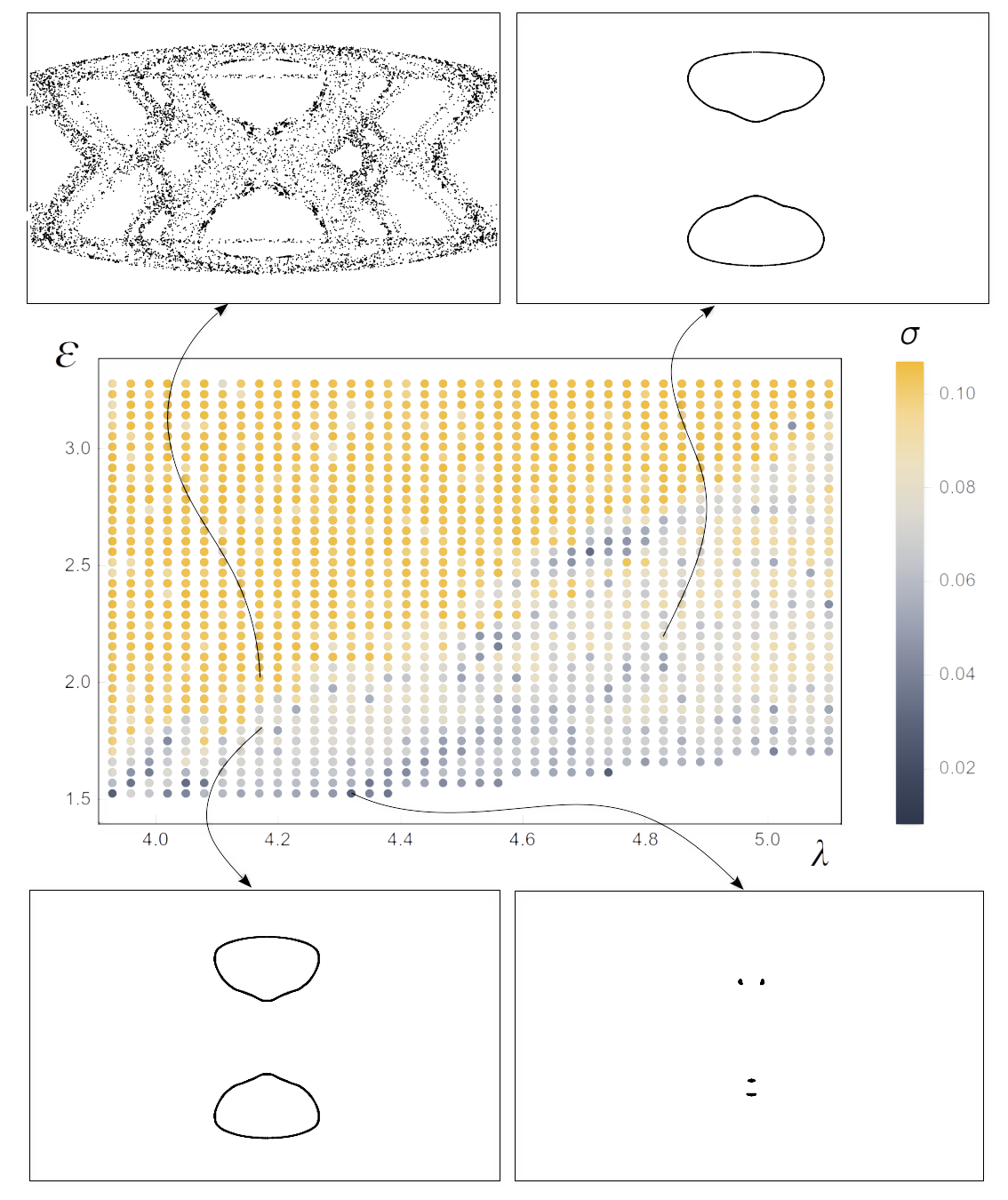}
\end{center}
\caption{Phase diagram in the energy ${\cal E}$ versus coupling $\lambda$ plane,   with non-vanishing vacuum expectation value. The initial magnetic flux vanishes $G(0)W(0)^2=0$ and the initial deviation from a purely Abelian configuration is $\dot W(0)=0.3818$. At each point, the chaotic nature of the solution is quantified by the mean quadratic dispersion of the points on the Poincaré section corresponding to the $\dot W$ versus $W$  plane. Some of such sections are depicted in the in-plots. The exact solution  \eqref{eq:solution.bothW.vev}-\eqref{eq:solution.bothG.vev} is now a single point in the phase diagram.}
\label{fig:phase.diagram.vev}
\end{figure}

\newpage

\begin{figure}[ht]
\begin{center}
\includegraphics[width=1\textwidth]{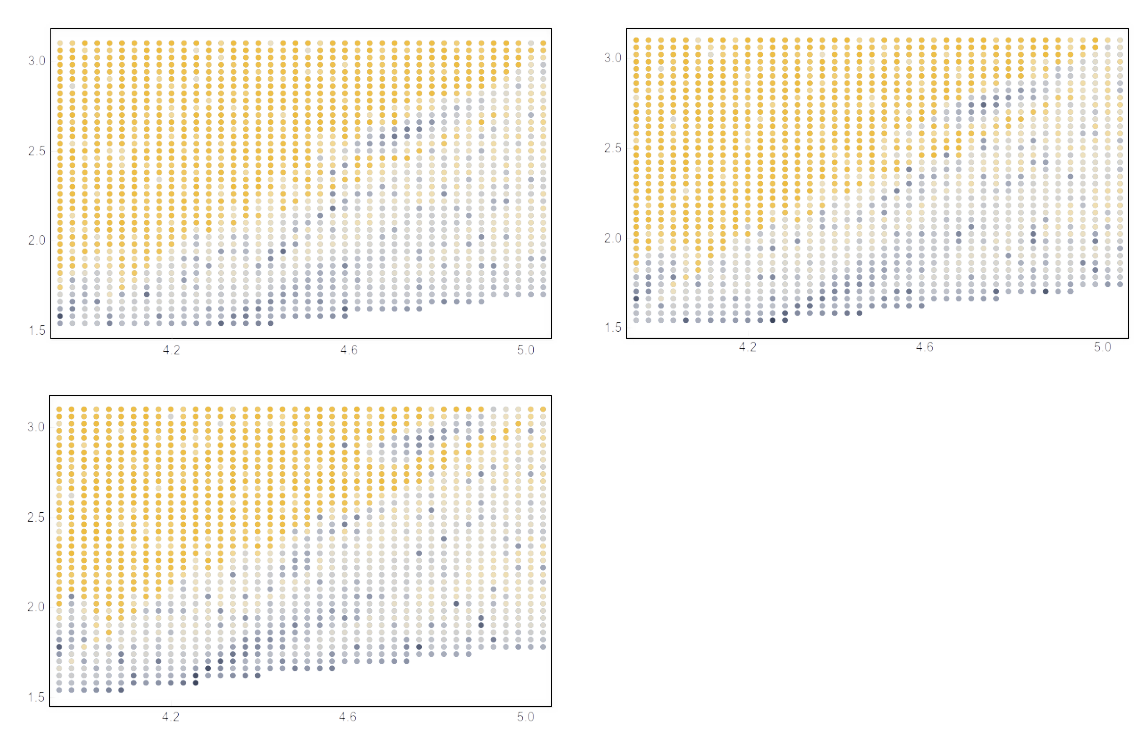}
\end{center} 
\vspace{-0.7cm}
\caption{Phase diagrams in the energy ${\cal E}$ versus coupling $\lambda$ plane, with non-vanishing vacuum expectation value, for different values of the initial deviation from a purely Abelian configuration $\dot W(0)$. The top row corresponds to $\dot W(0)=0.926515$ and $\dot W(0)=0.966515$ from left to right, while the bottom row has $\dot W(0)=1.01652$.  The initial magnetic flux vanishes $G(0) W(0)^2=0$.  }
\label{fig:phase.diagram.NA.vev}
\vspace{.7cm}
\end{figure}

\begin{figure}
\begin{center}
\includegraphics[width=1.0\textwidth]{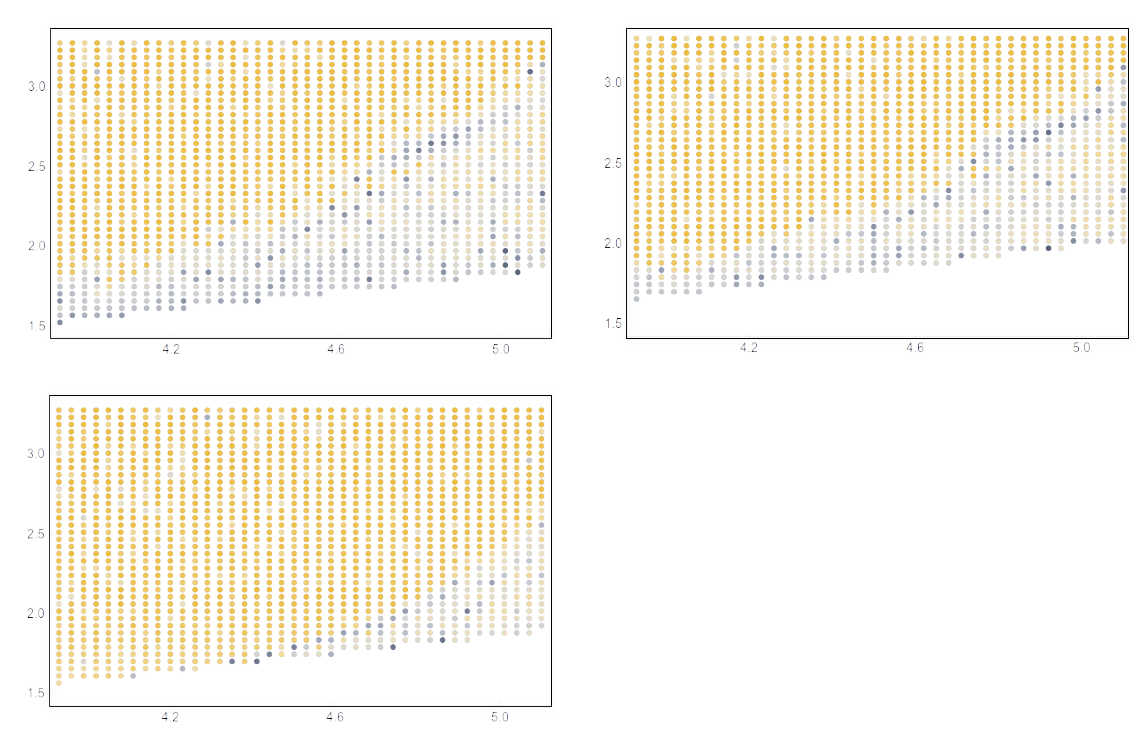}
\end{center}
\vspace{-0.5cm}
\caption{Phase diagrams in the energy ${\cal E}$ versus coupling $\lambda$ plane, with non-vanishing expectation value, for different values of the initial magnetic flux $G(0)W(0)^2$. The top row corresponds to $G(0)W(0)^2=0.028684$ and $G(0)W(0)^2=0.057368$ from left to right, while the middle row has $G(0)W(0)^2=0.14342$ and $G(0)W(0)^2=2^3$ respectively.  The initial deviation from a purely Abelian configuration is $\dot W(0)=0.916515$. }
\vspace{.7cm}
\label{fig:phase.diagram.flux.vev}
\end{figure}

The results for the Poincaré sections, the spectrum, and the geodesic deviation, are presented in Fig. \ref{fig:plots chaos con vev}. Again, as the energy grows, the system transitions to a chaotic regime. 

In Fig. \ref{fig:phase.diagram.vev} we can see the phase diagram, together with some Poincaré sections in the $\dot W$ versus $W$ plane. In this case, since the exact solution \eqref{eq:solution.bothG.vev}-\eqref{eq:solution.bothW.vev} has no constants of integration, its locus intersects the ${\cal E}$ versus $\lambda$ plane is a single point. Thus there is no stable ``bridge'' identifiable in the plot. 

Even if, due to the absence of the stability bridge of the vanishing vacuum expectation value case, the diagrams are more fuzzy, the general trends persist. Indeed, as the initial non-Abelian component of the configuration $\dot W(0)$ grows, the phase diagram is more regular, see Fig.\ref{fig:phase.diagram.NA.vev}. Conversely, as the initial magnetic flux $G(0)W(0)^2$ gets larger, the chaotic region gets wider \ref{fig:phase.diagram.flux.vev}. 

\newpage
\section{Probe scalar field}
\label{sec:resonance}

As the energy ${\cal E}$ of our mechanical system is conserved, we know that there is no radiation. Therefore, we may conclude that the configuration has no physical effects outside the coil, and thus the transition to chaos could be ``undetectable from the outside''. For that reason, in this section we studied its resonance effects on a probe scalar field presenting a possible mechanism to detect such transition.

In order to explore the features of these configurations let us consider a probe scalar field $\psi $ which transforms in the fundamental representation of $SU( 2)$. The covariant derivative is defined as
\begin{equation}
D_{\mu }\psi =\partial _{\mu }\psi +A_{\mu }\psi \ ,
\end{equation}
in such a way that %,as usual, the
its
commutator gives the field strength (\ref{eq:model.gaugecurvature}) \emph{i.e} $\left[ D_{\mu },D_{\nu }\right] \psi =F_{\mu \nu }\psi $. The action principle for the scalar field  is given by 
\begin{equation}
I[ \psi ,\psi ^{\dagger }] =-\int d^{4}x\sqrt{-g}\left( D_{\mu
}\psi \right) ^{\dagger }D^{\mu }\psi \ .  \label{action psi prueba}
\end{equation}
The equation coming from the variation of 
%the
this
action %principle 
(\ref{action psi prueba}) in the background  (\ref{eq:ansatz.YM}) expands as
\begin{eqnarray}
&&-\partial _{t}^{2}\psi +\partial _{z}^{2}\psi +\frac{1}{\rho }\partial
_{\rho }\psi +\partial _{\rho }^{2}\psi +\frac{1}{\rho ^{2}}\partial
_{\varphi }^{2}\psi -W^{2}\psi -\frac{1}{4\rho ^{2}}\psi + \label{eqpsi2} \\
&&\qquad\qquad\qquad\qquad\qquad
+\frac{2}{\rho ^{2}}\left( -\frac{W}{\sqrt{2}}\rho t_{1}-\frac{1}{2}%
t_{3}\right) \partial _{\varphi }\psi +\frac{1}{\rho }\frac{W}{\sqrt{2}}%
t_{2}\psi +\frac{2W}{\sqrt{2}}t_{2}\partial _{\rho }\psi \left. =\right. 0\ .
\notag
\end{eqnarray}
In order to apply time-dependent perturbation theory, to evaluate transitions amplitudes of the state of the scalar triggered by the interaction with the background %gauge 
field,
%$W(t)$, 
it is convenient to separate the above equations into two terms. The first term corresponds to
\begin{eqnarray}
 &&H_{0}\psi =-\partial _{t}^{2}\psi +\partial _{z}^{2}\psi +\frac{1}{\rho }%
\partial _{\rho }\psi +\partial _{\rho }^{2}\psi +\frac{1}{\rho ^{2}}%
\partial _{\varphi }^{2}\psi -\frac{1}{4\rho ^{2}}\psi   -\frac{t_{3}}{\rho ^{2}}\partial _{\varphi }\psi \left. \right. \  
\notag
\end{eqnarray}%
which defines the action of $H_0$ on the scalar, while the second term defines $H_{\sf int}$ by
\begin{equation}
H_{\sf int}\psi =-W^{2}\psi -\frac{\sqrt{2}W}{\rho ^{2}}t_{1}\partial _{\varphi
}\psi +\frac{1}{\rho }\frac{W}{\sqrt{2}}t_{2}\psi +\frac{2W}{\sqrt{2}}%
t_{2}\partial _{\rho }\psi  \ .
\end{equation}
This splitting allows to analyze the time-dependent part of the gauge field with time-dependent perturbation theory taking advantage of the fact that the ``unperturbed Hamiltonian'' $H_{0}\psi =0$ can be solved exactly. Hereafter we proceed in a canonical fashion, and
details can be found in the Appendix. 

Using the symbols $\uparrow,\downarrow$ to denote the up and down components of the field $\psi$, and the indices $n,\ell,m$ to identify the longitudinal, radial and angular modes respectively, we obtain the eigenstates of the free hamitonian $H_0$ as $|\uparrow n \ell m \pm\rangle$ and   $|\downarrow n \ell m \pm\rangle$ where $\pm$ denote left and right movers in the angular direction. The referred transition amplitude turns out to be given by
\small
\begin{eqnarray}
\langle \downarrow\ell' m'n'+|H_{\sf int}|\uparrow\ell mn-\rangle
&=&
%=
-\frac{\pi L%
\mathcal{N}_{\ell ^{\prime }m^{\prime }n^{\prime }}\mathcal{\bar{N}}_{\ell
mn}}{R_{0}\sqrt{2\omega _{\ell ^{\prime }m^{\prime }n^{\prime }}\bar{\omega}%
_{\ell mn}}}\delta _{m}^{m^{\prime }}\delta _{\ell }^{\ell ^{\prime }}\alpha
_{n}^{m-\frac{1}{2}} 
%\times \\ &&  \times
\int dt\,W( t) e^{i\left( \omega _{\ell^{\prime }m^{\prime }n^{\prime }}-\bar{\omega}_{\ell mn}\right) t} 
\times \nonumber \\ &&  \ \ \times
%\notag  
  \int_{0}^{R_{0}}d\rho\, \rho\, J_{m^{\prime }+\frac{1}{2}}\left( \alpha
_{n^{\prime }}^{m^{\prime }+\frac{1}{2}}\frac{\rho }{R_{0}}\right) J_{m+%
\frac{1}{2}}\left( \alpha _{n}^{m-\frac{1}{2}}\frac{\rho }{R_{0}}\right) \ ,\label{amplitude1}
\end{eqnarray}
\normalsize
where $J_{\eta }$ are Bessel functions and their zeros are labeled
by $\alpha _{n}^{\eta }$ with $n=1,2,\dots \ $, 
the constants ${\cal N}_{\ell mn}$ are for normalization in a cylinder of length $L$ and radius $R_0$, and $\omega_{\ell mn}$ denote the eigenfrequencies of the unperturbed Hamiltonian $H_0$.

The above formula for the transition amplitude
%corresponding
corresponds
to a probe scalar field, coupled to the time-dependent topologically non-trivial Yang-Mills-Higgs background, 
and it
is the main technical result of the present section. In particular, Eq. (\ref{amplitude1}) shows that if the classical background is in
its
%the
integrable phase, then as it has been discussed in the analysis of the Poincaré sections, the Fourier spectrum of the gauge field $W(t)$ has few relevant peaks. In these cases, the
transition
amplitude % $\Omega (\ell ^{\prime },m^{\prime },n^{\prime };\ \ell ,m,n)$
will be different from zero in such few cases, 
corresponding to the resonances between $\omega _{\ell ^{\prime }m^{\prime }n^{\prime }}-\bar{%
\omega}_{\ell mn}$ and the Fourier components of $W$. On the other hand, in the chaotic regime, the amplitude %\Omega (\ell ^{\prime },m^{\prime },n^{\prime };\ \ell ,m,n)$ 
will be different from zero in a broad band of
values for
$\omega _{\ell ^{\prime }m^{\prime }n^{\prime }}-\bar{\omega}_{\ell mn}$. Therefore, the
transition amplitudes of the probe scalar field can detect whether the non-Abelian coil is in the chaotic or integrable regime.

\section{Conclusions}

In the present paper we have discussed how the chaotic behavior of time-dependent configurations in the $SU(2)$ Georgi-Glashow model is affected by the Higgs coupling constant, by the vacuum expectation value as well as by the presence of topologically non-trivial fluxes, which in the present case
correspond to the flux of the non-Abelian magnetic field projected along the Higgs field. 

There are many intriguing questions which have not been analyzed in detail so far in the  literature. For instance: does
the presence of the Higgs potential and of the vacuum expectation value increase or decrease the chaotic behavior of the theory? What is the effect of non-trivial topological fluxes? The main problem to solve in order to answer them is related to the construction of a suitable Ansatz for the
gauge and Higgs fields. Indeed, one can easily write down explicit expressions both for the gauge and for the Higgs fields where all the
components depend on time only, as it is usually done in the literature on the chaotic behavior of Yang-Mills-Higgs theory: see \cite{chaosYM1}-\cite{chaosYMreview} and references therein. In this way the field equations reduce consistently to a dynamical system which can be analyzed using the known
tools of chaotic dynamics. However, if all the fields only depend on time, then the topological fluxes may vanish. This is the reason why it is useful to design an Ansatz in such a way that the fields depend in a non-trivial way also on the spatial coordinates, keeping alive the topological fluxes, but % in such a way that
with
the field equations 
%reduce
reducing
to an autonomous dynamical system. In the present work we have constructed such an Ansatz. 

As a general result, the transition to chaos occurs as the energy gets larger at fixed coupling. Moreover, as the initial magnetic flux is increased, the chaotic region of the phase diagram gets larger. Finally, the deviation of the solution from a purely Abelian configuration contributes to stability. These two last behaviours are somewhat intriguing, since on the one hand one would expect that topological fluxes stabilize the system, and on the other the non-linear character of the equations which is responsible of their chaotic nature, is inherited from the non-Abelianity of the theory.

A byproduct of the analysis is that we have also have identified an integrable sector where the field equations can be integrated analytically and the corresponding exact solutions represent the non-Abelian version of self-sustained alternating current generator. Moreover, the Ansatz has been constructed in such a way that one can, for instance, increase (or decrease) the control parameters, such as the Higgs coupling and the vacuum expectation value, and analyze how this change affects the chaotic properties. 
% of the system.
This situation is especially suitable to be studied using the tools introduced by Casetti, Pettini and Cohen (see \cite{cerruti-pettini-cohen} and references therein) in their geometric approach to the search for the stochasticity threshold in Hamiltonian dynamics. Using these tools we have shown that as one increases the energy, integrability is lost. Moreover, we
proved 
%have shown
that the chaotic behavior and sensitive dependence on the initial condition shown by the exponential growth of the geodesic deviation in Jacobi metric, are triggered by a parametric resonance.
Finally, we checked the transition to chaos can be observed also on the effective $U(1)$ field \eqref{eq:ansatz.Feef}. This is interesting since such a non-linear phenomenon is not expected in the standard linear $U(1)$ gauge dynamics, representing a genuine non-Abelian effect. 

\newpage

\subsection*{Acknowledgements}
F. C. and J. O. have been funded by Fondecyt Grants 1240048 and 1221504. The work of
M. O. is partially funded by Beca ANID de Doctorado 21222264.
N. G. wants to thank Centro de Estudios Científicos (CECs) and Universidad de Concepción (UdeC) by hospitality and support during this work. 
The Centro de Estudios Cient\'{\i}ficos (CECs) is funded by the Chilean Government through the Centers of Excellence Base Financing Program of ANID.

\appendix

\section{Perturbation theory}

The free Hamiltonian $H_{0}$, %given by (\ref{free hamiltonian}),
can be diagonalized by the following field configuration that fulfills the equation
(\ref{eqpsi2}) with $W=0$%
\begin{equation}
\Phi \left( t,\vec{x}\right) =\bar{\phi}\left( t,\vec{x}\right) +\phi \left(
t,\vec{x}\right) \ ,\label{the_expansion_Phi}
\end{equation}%
\small
\begin{eqnarray*}
\bar{\phi}^{a}\!\left( t,\vec{x}\right)\! &\equiv &\!\sum_{\ell mn}^{\text{%
{\LARGE \_}}}\!\frac{\mathcal{\bar{N}}_{\ell mn}}{\sqrt{2\bar{\omega}_{\ell mn}%
}}\sin\! \left( \frac{\pi \ell z}{L}\right) J_{m-\frac{1}{2}}\!\!\left( \frac{\alpha^{m-\frac{1}{2}}_{n}\rho }{R_{0}}\right) \!\left( 
\begin{array}{c}
\boldsymbol{l}_{\uparrow \ell mn}e^{i\left( m\varphi -\bar{\omega}_{\ell mn}t\right) }+%
\tilde{\boldsymbol{l}}_{\uparrow \ell mn}^{\ \dagger }e^{i\left( m\varphi +\bar{\omega}%
_{\ell mn}t\right) } \\ \!
\boldsymbol{l}_{\downarrow \ell mn}e^{-i\left( m\varphi +\bar{\omega}_{\ell mn}t\right) }+%
\tilde{\boldsymbol{l}}_{\downarrow \ell mn}^{\ \dagger }e^{-i\left( m\varphi -\bar{\omega}%
_{\ell mn}t\right)\! } \label{the_expansion}
\end{array} \!
\right) ^{\!\!a}  \\
\phi ^{a}\!\left( t,\vec{x}\right) \!&\equiv \!&\sum_{\ell mn}\!\frac{\mathcal{N}%
_{\ell mn}}{\sqrt{2\omega _{\ell mn}}}\sin\! \left( \frac{\pi \ell z}{L}\right)
J_{m+\frac{1}{2}}\!\!\left( \frac{\alpha^{m+\frac{1}{2}}_{n}\rho }{R_{0}}\right) \!\left( 
\begin{array}{c}\!
\boldsymbol{r}_{\uparrow \ell mn}e^{-i\left( m\varphi +\omega _{\ell mn}t\right) }+\tilde{%
\boldsymbol{r}}_{\uparrow \ell mn}^{\ \dagger }e^{-i\left( m\varphi -\omega _{\ell
mn}t\right) }\! \\ 
\boldsymbol{r}_{\downarrow \ell mn}e^{i\left( m\varphi -\omega _{\ell mn}t\right) }+%
\tilde{\boldsymbol{r}}_{\downarrow \ell mn}^{\ \dagger }e^{i\left( m\varphi +\omega
_{\ell mn}^{\prime }t\right) }%
\end{array}\!
\right) ^{\!\!a} 
\end{eqnarray*}
\normalsize
Where $J_\eta$ are the Bessel functions and $\alpha^\eta_n$ is the $n^{\text{th}}$ zero of the Bessel function $J_\eta$. In this expansion we have eight types of creation and annihilation operators $( \boldsymbol{l}_{\uparrow },\tilde{\boldsymbol{l}}_{\uparrow },\boldsymbol{l}_{\downarrow },\tilde{\boldsymbol{l}}_{\downarrow },\boldsymbol{r}_{\uparrow },$ $\tilde{\boldsymbol{r}}_{\uparrow },\boldsymbol{r}_{\downarrow },\tilde{\boldsymbol{r}}_{\downarrow })$. The states that we used in \eqref{amplitude1} are 
\begin{eqnarray}
   |\uparrow \ell m n - \rangle=\boldsymbol{l}_{\uparrow \ell m n}^{\dagger}|0\rangle \;, \qquad |\downarrow \ell' m' n' + \rangle=\boldsymbol{r}_{\downarrow \ell' m' n'}^{\dagger}|0\rangle \; ,
\end{eqnarray}
where $|0\rangle$ stands for the vacuum state of the theory. This solution satisfies the boundary conditions $\Phi |_{\partial M}=0$ where $%
M$ is the solid cylinder of radius $R_{0}$ and length $L$. The range of the integer numbers $\ell ,\ m,\ n$, as well as the definition of the summations in \eqref{the_expansion_Phi} are
\begin{equation}
\sum_{\ell mn}^{\text{{\LARGE \_}}}\equiv \sum_{\ell =1}^{\infty
}\sum_{m=1}^{\infty }\sum_{n=1}^{\infty }\ ,\qquad \qquad \sum_{\ell
mn}\equiv \sum_{\ell =1}^{\infty }\sum_{m=0}^{\infty }\sum_{n=1}^{\infty }\ .
\end{equation}%
The normalization constants are
\begin{equation}
\mathcal{\bar{N}}_{\ell mn}=\sqrt{\frac{2}{\pi L}}\frac{1}{R_{0}J_{m+\frac{1%
}{2}}\left( \alpha^{m-\frac{1}{2}}_{n}\right) }\ ,\qquad \quad \mathcal{N}_{\ell mn}=%
\sqrt{\frac{2}{\pi L}}\frac{1}{R_{0}J_{m+\frac{3}{2}}\left( \alpha^{m+\frac{1}{2}}_{n}\right) }\ ,
\end{equation}%
and the frequencies that are fixed by the boundary conditions are given by%
\begin{equation}
\left( \bar{\omega}_{\ell mn}\right) ^{2}=\left( \frac{\alpha^{m-\frac{1}{2}}_{n}}{%
R_{0}}\right) ^{2}+\left( \frac{\pi \ell }{L}\right) ^{2}\ ,\qquad \qquad
\left( \omega _{\ell mn}\right) ^{2}=\left( \frac{\alpha^{m+\frac{1}{2}}_{n}}{R_{0}}%
\right) ^{2}+\left( \frac{\pi \ell }{L}\right) ^{2}\ .
\end{equation}
The state with smallest energy in the system is given by $\bar{\omega}_{011}=\omega_{001}=(\alpha_1^\frac{1}{2}/R_0)^2$.

The conjugate momenta for the Lagrangian defined in (\ref{action psi prueba}), following the standard definitions are
\begin{eqnarray}
P_{a} &=&\sqrt{\gamma }\left( \partial _{t}\psi ^{\dagger }- \psi^{\dagger }A_{t}\right)_{a} \ , \\
P^{\prime a} &=&\sqrt{\gamma }\left( \partial _{t}\psi + A_{t}\psi \right)^{a} \ ,
\end{eqnarray}
here $\gamma $ is the determinant of the induced metric $\gamma _{\mu \nu }=g_{\mu \nu }+\delta _{\mu }^{t}\delta _{\nu }^{t}$, which is the spatial section of the metric (\ref{eq:ansatz.metric}), then $\sqrt{\gamma}=\rho$. The canonical momenta given in terms of \eqref{the_expansion_Phi} through the above definition, forms a representation of the canonical algebra
\begin{equation}
\left[ \Phi ^{a}\left( t,\mathbf{x}\right) ,P_{b}\left( t,\mathbf{y}\right) %
\right] =i \delta _{b}^{a}\delta \left( \mathbf{x}-\mathbf{y}\right) \
,\quad \left[ \Phi _{a}^{\dagger }\left( t,\mathbf{x}\right) ,P'^{b}\left( t,%
\mathbf{y}\right) \right] =i \delta _{a}^{b}\delta \left( \mathbf{x}-%
\mathbf{y}\right) \ ,
\end{equation}%
with the following commutation relation for the creation/annihilation operators 
\small
\begin{eqnarray*}
\left[ \boldsymbol{l}_{\uparrow \ell mn},\boldsymbol{l}_{\uparrow \ell ^{\prime }m^{\prime }n^{\prime
}}^{\dagger }\right]  &=&\left[ \tilde{\boldsymbol{l}}_{\uparrow \ell mn},\tilde{\boldsymbol{l}}%
_{\uparrow \ell ^{\prime }m^{\prime }n^{\prime }}^{\dagger }\right] =\left[
\boldsymbol{r}_{\uparrow \ell mn},\boldsymbol{r}_{\uparrow \ell ^{\prime }m^{\prime }n^{\prime
}}^{\dagger }\right] =\left[ \tilde{\boldsymbol{r}}_{\uparrow \ell mn},\tilde{\boldsymbol{r}}%
_{\uparrow \ell ^{\prime }m^{\prime }n^{\prime }}^{\dagger }\right] =\delta
_{\ell }^{\ell ^{\prime }}\delta _{m}^{m^{\prime }}\delta _{n}^{n^{\prime
}}\ , \\
\left[ \boldsymbol{l}_{\downarrow \ell mn},\boldsymbol{l}_{\downarrow \ell ^{\prime
}m^{\prime }n^{\prime }}^{\dagger }\right] &=&\left[ \tilde{\boldsymbol{l}}_{\downarrow
\ell mn},\tilde{\boldsymbol{l}}_{\downarrow \ell ^{\prime }m^{\prime
}n^{\prime }}^{\dagger }\right] =\left[ \boldsymbol{r}_{\downarrow \ell mn},\boldsymbol{r}_{\downarrow \ell ^{\prime }m^{\prime
}n^{\prime }}^{\dagger }\right]  =\left[ \tilde{\boldsymbol{r}}_{\downarrow \ell mn},%
\tilde{\boldsymbol{r}}_{\downarrow \ell ^{\prime }m^{\prime }n^{\prime }}^{\dagger }%
\right] =\delta _{\ell }^{\ell ^{\prime }}\delta
_{m}^{m^{\prime }}\delta _{n}^{n^{\prime }}\ .
\end{eqnarray*}
\normalsize

To see how this works, we compute one commutator between $\Phi ^{a}$ and $%
P_{b}$. The following representations of the Dirac delta will be useful%
\begin{eqnarray}
\delta \left( \varphi -\varphi ^{\prime }\right)  &=&\sum_{m=-\infty
}^{\infty }\frac{1}{2\pi }e^{im\left( \varphi -\varphi ^{\prime }\right) }\ ,
\nonumber \\
\delta \left( \rho -\rho ^{\prime }\right)  &=&\sum_{n=1}^{\infty }\frac{%
2\rho }{R_{0}^{2}J_{\eta +1}\left( \alpha _{n}^{\eta }\right) ^{2}}J_{\eta
}\left( \frac{\alpha _{n}^{\eta }\rho }{R_{0}}\right) J_{\eta }\left( \frac{%
\alpha _{n}^{\eta }\rho ^{\prime }}{R_{0}}\right) \ , \\
\delta \left( z-z^{\prime }\right)  &=&\sum_{\ell =1}^{\infty }\frac{2}{L}%
\sin \left( \frac{\pi \ell z^{\prime }}{L}\right) \sin \left( \frac{\pi \ell
z}{L}\right) \ .  \nonumber
\end{eqnarray}
The expression of $P_{b}$ following the definition for our magnetic background  is%
\begin{equation}
P_{a}\equiv \rho\partial _{t}\Phi ^{\dagger }\equiv \bar{p}%
_{a}+p_{a}\ ,
\end{equation}
where
\small
\begin{eqnarray}\nonumber
\bar{p}_{a}\! &=&\!\sum_{\ell mn}^{\text{{\LARGE \_}}}\rho i\mathcal{%
\bar{N}}_{\ell mn}\sqrt{\frac{\bar{\omega}_{\ell mn}}{2}}\sin \!\left( \frac{%
\pi \ell z}{L}\right) J_{m-\frac{1}{2}}\!\left( \frac{\bar{\chi}_{n}^{m}\rho }{%
R_{0}}\right) \!\left( 
\begin{array}{c}
\boldsymbol{l}_{\uparrow \ell mn}^{\dagger }e^{-i\left( m\varphi -\bar{\omega}_{\ell mn}t\right) }-%
\tilde{\boldsymbol{l}}_{\uparrow \ell mn}e^{-i\left( m\varphi +\bar{\omega}_{\ell mn}t\right) } \\ 
\boldsymbol{l}_{\downarrow \ell mn}^{\dagger }e^{i\left( m\varphi +\bar{\omega}_{\ell mn}t\right) }-%
\tilde{\boldsymbol{l}}_{\downarrow \ell mn}e^{i\left( m\varphi -\bar{\omega}_{\ell mn}t\right) }%
\end{array}%
\right) ^{\!\!T} , \\
p_{a}\! &=&\!\sum_{\ell mn}\rho i\mathcal{N}_{\ell mn}\sqrt{\frac{%
\omega _{\ell mn}}{2}}\sin\! \left( \frac{\pi \ell z}{L}\right) J_{m+\frac{1}{2%
}}\!\left( \frac{\chi _{n}^{m}\rho }{R_{0}}\right) \!\left( 
\begin{array}{c}
\boldsymbol{r}_{\uparrow\ell mn}^{\dagger }e^{i\left( m\varphi +\omega _{\ell mn}t\right) }-%
\tilde{\boldsymbol{r}}_{\uparrow \ell mn}e^{i\left( m\varphi -\omega _{\ell mn}t\right) } \\ 
\boldsymbol{r}_{\downarrow \ell mn}^{\dagger }e^{-i\left( m\varphi -\omega _{\ell mn}t\right) }-%
\tilde{\boldsymbol{r}}_{\downarrow \ell mn}e^{-i\left( m\varphi +\omega _{\ell mn}t\right) }%
\end{array}%
\right) ^{\!\!T},  \nonumber\\
\bar{\chi}_{n}^{m}&\equiv& \alpha_{n}^{m-1}\; , \quad \chi_{n}^{m}\equiv \alpha_{n}^{m+1} \; .\nonumber
\end{eqnarray}%
\normalsize
The commutator can be written as%
\begin{equation}
\left[ \Phi ^{a}\left( t,\vec{x}\right) ,P_{b}\left( t,\vec{x}^{\prime
}\right) \right] =\left( \left[ \bar{\phi}^{1},\bar{p}_{1}\right] +\left[
\phi ^{1},p_{1}\right] \right) \delta _{1}^{a}\delta _{b}^{1}+\left( \left[ 
\bar{\phi}^{2},\bar{p}_{2}\right] +\left[ \phi ^{2},p_{2}\right] \right)
\delta _{2}^{a}\delta _{b}^{2}\ .  \label{phi P}
\end{equation}%
Let us compute explicitly the first parenthesis of \eqref{phi P}. The first commutator is%
\begin{eqnarray}
\left[ \bar{\phi}^{1},\bar{p}_{1}\right]  &=&\sum_{\ell mn}^{\text{{\LARGE %
\_}}}i\rho \mathcal{\bar{N}}_{\ell mn}^{2}\sin \left( \frac{\pi \ell z}{L}%
\right) \sin \left( \frac{\pi \ell z^{\prime }}{L}\right) J_{m-\frac{1}{2}%
}\left( \frac{\bar{\chi}_{n}^{m}\rho }{R_{0}}\right) J_{m-\frac{1}{2}}\left( 
\frac{\bar{\chi}_{n}^{m}\rho ^{\prime }}{R_{0}}\right) e^{im\left( \varphi
-\varphi ^{\prime }\right) }\ ,  \nonumber \\
&=&\delta \left( z-z^{\prime }\right) \sum_{\ell mn}^{\text{{\LARGE \_}}}i
\frac{1}{2\pi }\frac{2}{R_{0}^{2}J_{m+\frac{1}{2}}\left( \bar{\chi}%
_{n}^{m}\right) ^{2}}\rho J_{m-\frac{1}{2}}\left( \frac{\bar{\chi}%
_{n}^{m}\rho }{R_{0}}\right) J_{m-\frac{1}{2}}\left( \frac{\bar{\chi}%
_{n}^{m}\rho ^{\prime }}{R_{0}}\right) e^{im\left( \varphi -\varphi ^{\prime
}\right) }\ ,  \nonumber \\
&=&i\frac{1}{2\pi }\delta \left( z-z^{\prime }\right) \delta \left( \rho
-\rho ^{\prime }\right) \sum_{m=1}^{\infty }e^{im\left( \varphi -\varphi
^{\prime }\right) }\ .
\end{eqnarray}%
while the second term is
\small
\begin{eqnarray}
\left[ \phi ^{1},p_{1}\right]  
\!&=&\!\sum_{\ell mn}\frac{2i}{\pi L}\frac{\rho}{R_{0}^{2}J_{m+\frac{3}{2}%
}\!\left( \chi _{n}^{m}\right) ^{2}}\sin\! \left( \frac{\pi \ell z}{L}\right)
\sin\! \left( \frac{\pi \ell z^{\prime }}{L}\right) J_{m+\frac{1}{2}}\!\left( 
\frac{\chi _{n}^{m}\rho }{R_{0}}\right) J_{m+\frac{1}{2}}\!\left( \frac{\chi
_{n}^{m}\rho ^{\prime }}{R_{0}}\right) e^{-im\left( \varphi -\varphi
^{\prime }\right) }\ ,  \nonumber \\
&=&\sum_{m=0}^{\infty }i\frac{1}{2\pi }\delta \left( z-z^{\prime }\right)
\delta \left( \rho -\rho ^{\prime }\right) e^{-im\left( \varphi -\varphi
^{\prime }\right) }\ .
\end{eqnarray}%
\normalsize
Replacing in the first parenthesis in (\ref{phi P})%
\begin{equation}
\left[ \bar{\phi}^{1},\bar{p}_{1}\right] +\left[ \phi ^{1},p_{1}\right]
=i\delta \left( z-z^{\prime }\right) \delta \left( \rho -\rho ^{\prime
}\right) \frac{1}{2\pi }\left( \sum_{m=1}^{+\infty }e^{im\left( \varphi
-\varphi ^{\prime }\right) }+\sum_{m=0}^{\infty }e^{-im\left( \varphi
-\varphi ^{\prime }\right) }\right) \ ,
\end{equation}%
changing the sign in the second summation and using the fact that we get the representation of the delta function, thus
\begin{equation}
\left[ \bar{\phi}^{1},\bar{p}_{1}\right] +\left[ \phi ^{1},p_{1}\right]
=i\delta \left( z-z^{\prime }\right) \delta \left( \rho -\rho ^{\prime
}\right) \delta \left( \varphi -\varphi ^{\prime }\right) \ .
\end{equation}%
One can show that the same mechanism works for the second parenthesis in (\ref%
{phi P}),%
\begin{equation}
\left[ \bar{\phi}^{2},\bar{p}_{2}\right] +\left[ \phi ^{2},p_{2}\right]
=i\delta \left( z-z^{\prime }\right) \delta \left( \rho -\rho ^{\prime
}\right) \delta \left( \varphi -\varphi ^{\prime }\right) \ .
\end{equation}
Replacing back into (\ref{phi P}) we find
\begin{equation}
\left[ \Phi ^{a}\left( t,\vec{x}\right) ,P_{b}\left( t,\vec{x}^{\prime
}\right) \right] =i\delta _{b}^{a}\delta \left( z-z^{\prime }\right) \delta
\left( \rho -\rho ^{\prime }\right) \delta \left( \varphi -\varphi ^{\prime
}\right) \ ,
\end{equation}
as promised.

\end{document}